\documentclass[structabstract]{aa}  
\usepackage{longtable,lscape}
 \usepackage{graphicx}
\usepackage{natbib}
\bibpunct{(}{)}{;}{a}{}{,}
\usepackage{txfonts}

\begin{document}

   \title{Spectroscopy of new brown dwarf members of $\rho$~Ophiuchi and an updated initial mass function
   \thanks{Based on observations made at the ESO La Silla and Paranal Observatory under program 085.C-0960.}}
\author{C. Alves de Oliveira\inst{1}, E. Moraux\inst{1}, J. Bouvier\inst{1}, H. Bouy\inst{2}}
\institute{UJF-Grenoble 1 / CNRS-INSU, Institut de Plan\'etologie et d'Astrophysique de Grenoble (IPAG) UMR 5274, Grenoble, F-38041, France  \\
\and
Centro de Astrobiolog\'{\i}a (INTA-CSIC); LAEFF, P.O. Box 78, 28691 Villanueva de la Ca\~{n}ada, Spain \\}

\date{Received 10 October 2011 / Accepted 30 December 2011}

  \abstract
  {}
   {To investigate the universality hypothesis of the initial mass function in the substellar regime, the population of the $\rho$~Ophiuchi molecular cloud is analysed by including a new sample of low-mass spectroscopically confirmed members. To that end, we have conducted a large spectroscopic follow-up of young substellar candidates uncovered in our previous photometric survey.} 
{The spectral types and extinction were derived for a newly found population of substellar objects, and its masses estimated by comparison to evolutionary models. A thoroughly literature search was conducted to provide an up-to-date census of the cluster, which was then used to derive the luminosity and mass functions, as well as the ratio of brown dwarfs to stars in the cluster. These results were then compared to other young clusters.}
  {It is shown that the study of the substellar population of the $\rho$~Ophiuchi molecular cloud is hampered only by the high extinction in the cluster ruling out an apparent paucity of brown dwarfs. The discovery of 16 new members of $\rho$~Ophiuchi, 13 of them in the substellar regime, reveals the low-mass end of its population and shows the success of our photometric candidate selection with the WIRCam survey. The study of the brown dwarf population of the cluster reveals a high disk fraction of 76$^{+5}_{-8}$\%. }
   {Taking the characteristic peak mass of the derived mass function and the ratio of brown dwarfs to stars into account, we conclude that the mass function of $\rho$~Ophiuchi is similar to other nearby young clusters.}

\titlerunning{Spectroscopy of new brown dwarf members in $\rho$ Ophiuchi} 
\authorrunning{C. Alves de Oliveira et al.} 

   \maketitle

%________________________________________________________________

\section{Introduction}
\label{intro}

Although there is a generally accepted theory of how a star forms through the collapse of a molecular cloud core and contracts to the main sequence \citep[e.g.,][]{Shu1987}, two major breakthroughs in the mid-90s have added a new complexity to the star formation field: the discovery of exoplanets \citep{Mayor1995} and brown dwarfs \citep{Oppenheimer1995,Rebolo1995,Nakajima1995}. Brown dwarfs are a key element in understanding the origin of stellar masses and their distribution, the initial mass function (IMF), and in identifying the plethora of hospitable hosts where planetary systems can form. However, there is no current consensus regarding their formation mechanism and several scenarios are currently proposed. Some defend a star-like formation, i.e. fragmentation of molecular clouds into low-mass cores driven by turbulence \citep{Padoan2002,Hennebelle2008} or gravity \citep{Bonnell2008}. But distinct formation processes are also likely to take place, such as gravitational instabilities in disks \citep{Stamatellos2009}, premature ejection from prestellar cores \citep{Reipurth2001}, or photo-erosion of cores \citep{Kroupa2003}. The observational properties of brown dwarfs in young clusters show a global scaling-down trend from those of stars, arguing in favour of a common formation scenario \citep[e.g.,][]{Luhman2007}. However, most studies in young star-forming regions suffer from incompleteness both at lower masses (below 40~M$_{Jup}$) and in spatial content, frequently focusing on the inner regions of clusters. It is therefore unknown whether, as one moves to lower masses, other formation mechanisms take over \citep[e.g.,][]{Whitworth2010}. At the same time, the increasing capacity of numerical simulations to reproduce the collapse of entire molecular clouds is such that the comparison between simulations and observations starts to be hampered by our empirical knowledge based on small number statistics of the low-mass end of the brown dwarf regime \citep[e.g.,][]{Bate2009}. Extensive studies of the lowest mass populations in different environments and stages of evolution are therefore required to address these questions. 

To that end, several large photometric surveys are taking place to identify candidate brown dwarfs in various young clusters: for example, the Galactic Cluster Survey (GCS) \citep[e.g.,][]{Lodieu2011} part of the Infrared Deep Sky Survey \citep[UKIDSS,][]{Lawrence2007} at the UKIRT (United Kingdom Infrared Telescope), the SONYC survey (Substellar Objects in Nearby Young Clusters) \citep[e.g.][]{Scholz2009}, or our WIRCam/CFHT \citep[Wide Field IR Camera at the Canada France Hawaii Telescope,][]{Puget2004} survey of nearby star-forming regions \citep[e.g.][]{Burgess2009}, among other substellar studies of targeted clusters \citep[e.g.,][]{Bayo2011,Bejar2011}. Spectroscopic confirmation of photometric candidates in all the regions is needed, however, to construct an unbiased IMF, since the contamination by field dwarfs, giants, or even galaxies at the faint magnitude ranges being probed is known to be high. This is the motivation behind the observations presented in this work. From the initial list of 110 brown dwarf candidate members of $\rho$~Oph uncovered through the WIRCam survey \citep{AlvesdeOliveira2010}, we had confirmed spectroscopically seven new members in our previous pilot study, six of which were classified as brown dwarfs and one very low mass star. Here, we present the results of an extended spectroscopic follow-up of 34 additional targets. With this survey, we have observed spectroscopically 70\% of our candidates that fall within the photometric completeness limit of the WIRCam survey, and are nearly complete ($\sim$93\% of candidates observed) down to a few Jupiter masses, through a visual extinction of 20~mag.

In Sect.~\ref{obs} we describe the new spectroscopic observations and data reduction procedure. Section~\ref{method} explains the method adopted for deriving the spectral type and extinction for each candidate member, and the final results are presented in Sect.~\ref{results}. We discuss the properties of the newly found members and their comparison to the cluster's population through Sects.~\ref{discussion1} and \ref{discussion2}, and present the final conclusions in Sect. ~\ref{conclusion}.

%__________________________________________________________________

\section{Observations and data reduction} \label{obs}

\subsection{The ISAAC / VLT spectroscopic follow-up}
We have conducted a near-infrared (near-IR) spectroscopic follow-up of 34 targets from our WIRCam/CFHT list of candidate members of $\rho$~Oph, randomly selected with magnitudes brighter than the WIRCam survey completeness limits at \emph{J}$=$20.5 and \emph{H}$=$18.9~mag, and through a visual extinction of approximately 20~magnitudes (see Sect.~\ref{limitedsample} and Fig.~\ref{cmd}). We additionally observed five previously known brown dwarfs with spectral types determined in the optical, which we use to test our analysis methods (see Tables~\ref{log} and \ref{knownBD}). The observations were carried out in visitor mode from 4 to 6 May 2010 with ISAAC \citep[Infrared Spectrometer And Array Camera,][]{Moorwood1998} mounted on the Unit Telescope 3 (UT3) of the VLT, Paranal (Table~\ref{log}). Each target was observed with the low-resolution grating in the short wavelength arm of ISAAC and a long slit with a width of 0.6" or 0.8" to better match the seeing conditions. Given the faint \emph{J} magnitudes of most of the targets, observations were done only in the \emph{H} (1.42$-$1.82$\mu$m) and \emph{K} (1.82$-$2.5$\mu$m) bands. Individual exposure times were scaled to the target's brightness and night conditions, and repeated in an ABBA pattern to allow for sky subtraction. The slit position was aligned with the parallactic angle, except for targets CFHTWIR-Oph~20 and 98, which were aligned with a nearby faint target. Standard A0 stars were observed at regular intervals during the night.disk

The data was reduced with IRAF (Image Reduction and Analysis Facility) for all targets by following standard procedures. The 2D spectra were flat-fielded and corrected for bad pixels, after which the sky subtraction was done by subtracting adjacent pairs of spectra. The spectra were corrected for distortions along the spectral and spatial directions, wavelength-calibrated, and then aligned and co-added. Individual frames where the spectrum was strongly affected by bad pixels or cosmic rays were removed before the co-adding operation. The exposure times per individual integration and the final number of combined frames for each target are given in Table~\ref{log}. Each spectrum was extracted with the \emph{APALL} routine in IRAF. Each extracted spectrum was further divided by that of the standard A0 stars observed at similar airmass to remove the telluric absorptions, and the relative flux recovered by multiplying it by a theoretical spectrum of an A0 star \citep[taken from the ESO webpage]{Pickles1998}. A linear interpolation was made to remove the strong intrinsic hydrogen absorption lines from the spectra of the A0 standard stars,  across the lines that are more predominant at this resolution (the Brackett series lines at 1.54, 1.56, 1.57, 1.59, 1.61, 1.64, 1.68, 1.74, 2.17~$\mu$m). To calibrate the relative fluxes of the spectra originating from the two grisms, we calculated the pseudo spectral flux for each band by convolving the spectra with the respective WIRCam transmission curves and integrating the flux, which are then compared to the flux ratios for \emph{H} and \emph{K$_{s}$} band measurements of the WIRCam photometry, and scaled accordingly. 

\begin{table*}
\centering             
\caption{Journal of the observations.}            
\begin{tabular}{l c l l}        
\hline \hline
Date & Slit (") &  CFHTWIR-Oph target & Known brown dwarfs from literature \\
 \hline
 \multicolumn{4}{c}{ISAAC/VLT\tablefootmark{a}} \\
 \hline
2010 May 04 &	0.8   & 7 (120\emph{s}$\times$5, 120\emph{s}$\times$10)   & [L2007b]~Cha J11070768$-$7626326  (120\emph{s}$\times$14, 120\emph{s}$\times$10)\\
~   		&	~   & 9 (60\emph{s}$\times$9, 60\emph{s}$\times$10)   &  OTS44 (60\emph{s}$\times$10, 60\emph{s}$\times$8) \\
~   		&	~   & 12  (60\emph{s}$\times$9, 60\emph{s}$\times$10) &  ~  \\
~   		&	~   & 15  (40\emph{s}$\times$2, 20\emph{s}$\times$2) &  ~  \\
~   		&	~   & 16  (40\emph{s}$\times$2, 30\emph{s}$\times$2) &   ~ \\
~   		&	~   & 17  (60\emph{s}$\times$8, 60\emph{s}$\times$4) &   ~ \\
~   		&	~   & 31  (20\emph{s}$\times$2, 10\emph{s}$\times$2) &  ~  \\
~   		&	~   & 33  (120\emph{s}$\times$9, 120\emph{s}$\times$9) &  ~  \\
~   		&	~   & 66  (60\emph{s}$\times$10, 60\emph{s}$\times$8) &  ~  \\
~   		&	~   & 77  (60\emph{s}$\times$9, 60\emph{s}$\times$10) &  ~  \\
~   		&	~   & 90 (60\emph{s}$\times$8, 60\emph{s}$\times$4) &  ~  \\
~   		&	~   & 100  (60\emph{s}$\times$8, 60\emph{s}$\times$9) &  ~  \\
~   		&	~   & 103  (60\emph{s}$\times$9, 60\emph{s}$\times$10) &  ~  \\
2010 May 05   		&	0.6   & 52  (60\emph{s}$\times$2, 60\emph{s}$\times$2)  &   ~  \\
~ 		   		&	~   & 98    (60\emph{s}$\times$15, 60\emph{s}$\times$12)  &   ~  \\
~		   		&	0.8   & 19 (120\emph{s}$\times$12, 120\emph{s}$\times$10)    & CHSM17173  (60\emph{s}$\times$6, 60\emph{s}$\times$6)  \\
~		   		&	~   & 20    (120\emph{s}$\times$9, 120\emph{s}$\times$10)    & [NC98]~Cha~HA~11 (60\emph{s}$\times$7, 60\emph{s}$\times$8)  \\
~		   		&	~   & 30    (40\emph{s}$\times$2, 20\emph{s}$\times$2) 	  & ~   \\  
~		   		&	~   & 37    (40\emph{s}$\times$4, 20\emph{s}$\times$4) 	  & ~   \\  
~		   		&	~   & 42    (120\emph{s}$\times$16, 120\emph{s}$\times$14) 	  & ~   \\  
~		   		&	~   & 56   (60\emph{s}$\times$2, 60\emph{s}$\times$2) 	  & ~   \\  
~		   		&	~   & 69   (20\emph{s}$\times$2, 20\emph{s}$\times$2) 	  & ~   \\ 
~		   		&	~   & 78   (60\emph{s}$\times$4, 30\emph{s}$\times$4) 	  & ~   \\  
~		   		&	~   & 107   (20\emph{s}$\times$2, 20\emph{s}$\times$2) 	  & ~   \\  
2010 May 06   		&	0.6   & 1   (120\emph{s}$\times$14, 120\emph{s}$\times$7)   &   ~  \\
~		   		&	~    & 3   (60\emph{s}$\times$2, 60\emph{s}$\times$2)   &   ~  \\
~		   		&	~    & 6   (120\emph{s}$\times$8, 60\emph{s}$\times$5)   &   ~  \\
~		   		&	~    & 18   (120\emph{s}$\times$13, 120\emph{s}$\times$14)   &   ~  \\
~		   		&	~    & 22   (60\emph{s}$\times$4, 60\emph{s}$\times$4)   &   ~  \\
~		   		&	~    & 28   (120\emph{s}$\times$20, 60\emph{s}$\times$10)   &   ~  \\
~		   		&	~    & 53    (60\emph{s}$\times$7, 60\emph{s}$\times$4)   &   ~  \\
~		   		&	~    & 101  (120\emph{s}$\times$6, 60\emph{s}$\times$6)   &   ~  \\
~		   		&	~    & 104  (120\emph{s}$\times$6, 60\emph{s}$\times$6)   &   ~  \\
 ~		   		&	0.8   &  58 (60\emph{s}$\times$2, 60\emph{s}$\times$4) & TWA~26 (60\emph{s}$\times$8, 60\emph{s}$\times$8)    \\ 
 \hline
 \multicolumn{4}{c}{Osiris/GTC} \\
 \hline
2010 May 22   	&	1.0   & 31 (450\emph{s}$\times$2)  & ~ \\
2010 May 23   	&	1.0   & 30 (1500\emph{s}$\times$2) & ~ \\
~		   	&	~     & 96  (450\emph{s}$\times$2)  & ~ \\
\hline
\label{log} 
\end{tabular}

\tablefoottext{a}{Exposure times are given in seconds per individual integration, times the number of total integrations for the \emph{H} and \emph{K} grisms, respectively.}
\end{table*}

\subsection{The Osiris / GranTeCan observations}
We obtained optical spectra for three targets from the WIRCam survey, two of them were also observed in the near-IR during the ISAAC/VLT run previously described (CFHTWIR-Oph~30, 31), while the other had been observed with SofI/NTT and presented in our survey paper \citep[CFHTWIR-Oph~96]{AlvesdeOliveira2010}. The targets were observed with OSIRIS \citep{Cepa2000}, a spectrograph operating from 0.365 to 1.05~$\mu$m, mounted on the Gran Telescopio Canarias (GTC), the 10.4~m telescope at the Observatory del Roque de Los Muchachos (La Palma, Canary Islands). We obtained low-resolution spectroscopy for all targets with the R300R grism and an 1$\arcsec$ slit aligned with the parallactic angle. The observations were obtained in service mode on 22 and 23 May 2010, as well as the spectro-photometric standards, flats, bias, and arcs necessary for the calibrations. Each target was observed with two exposures of 450~s each for CFHTWIR-Oph~31 and 96, and 1500~s for CFHTWIR-Oph~30. The data was reduced using standard IRAF routines, including bias subtraction, flat-field correction, extraction of the 1D spectrum, and wavelength calibration. The instrumental response was corrected using spectrophotometric standards, and the individual spectra were median-combined. 

%__________________________________________________________________

\section{Methods} \label{method}
\subsection{Spectral classification in the near-IR}
It has been shown by several studies that low-resolution near-IR spectroscopy can be used to accurately determine spectral types of young stellar objects and confirm their low gravity \citep[e.g.,][]{Allers2007,Weights2009}. One of the most evident youth indicators is the water vapour absorption bands that result in the  triangular shape of the \emph{H}-band \citep{Luhman1999,Lucas2001,McGovern2004}, with a peak redwards of that of field dwarfs that instead show a plateau at the same wavelength range. The same effect of H$_{2}$O absorption is seen in the \emph{K} band ($\gtrsim$2.3$\mu$m), where CO absorption is also present. Provided the spectrum has a reasonable signal-to-noise ratio (S/N$\sim$30), the Na~I feature at 2.2$\mu$m can be used as a youth indicator, since it is less prominent in young objects than in field dwarfs. Finally, near-IR emission lines characteristic of outflows (e.g., H2 1$-$0 S(0) emission at 2.12~$\mu$m) or accretion (e.g., hydrogen recombination line Br$\gamma$ at 2.16~$\mu$m) can be detected even at low resolution and provide further evidence of the membership and youth of a candidate. In this section, we describe the methodology developed to investigate the nature of the candidate members of the $\rho$~Ophiuchi cluster.

\subsubsection{Empirical grid of template spectra}
To derive the spectral type and extinction of the candidate members observed with ISAAC, we compare each spectrum to an empirical grid of near-IR  template spectra of young stellar objects while varying the extinction. The objects that integrate the grid were chosen out of a list of the spectra of 37 confirmed members of Taurus and IC~348 (1-3Myr), which have an approximate age to that of $\rho$~Oph, and have spectral types from M3 to M9.5 classified in the optical through comparison to averages of dwarfs and giants \citep{Briceno2002,Luhman2003taurus,Luhman2003ic348,Luhman2004}. To determine the extinction towards each of the template spectra, the spectro-photometric colours were calculated by converting to magnitudes the integrated flux of each spectra convolved with the 2MASS transmission filter curves, and using the respective fluxes for zero magnitude \citep{Cohen2003}. The extinction (A$_{\emph{V}}$) was determined by comparing those to the expected \emph{J}$-$\emph{H} and \emph{H}$-$\emph{K} colours for young objects \citep[see Table~13]{Luhman2010}, and applied to deredden each spectrum accordingly. 

Based on their photometric properties and evolutionary models, the candidate members we observed could have spectral types as late as L5. Therefore, we added to this list near-IR spectra (K. Luhman, private communication) of optical standard young-field L dwarfs \citep{Cruz2009} with spectral types from L0 to L5 and spectral features indicative of low gravity. Although the membership of these objects to nearby loose associations has yet to be confirmed, the low-gravity features in their spectra are consistent with an age of $\approx$10Myr. Even taking the most conservative approach to this study, the behaviour of the gravity-sensitive indicators in their spectra (mainly from the alkali lines and the VO bands, see Kirkpatrick 2008) imply an upper-limit for their age of 100Myr making them more suitable for spectral classification of young brown dwarfs than old field dwarfs. Table~\ref{temp} lists the objects selected to build the grid of young templates, including their identification, together with the extinction values derived for the members of IC~348 and Taurus. 2MASS photometric errors (between 0.05 to 0.11~mag) are approximately the same as the differences between photometric and synthetic colours. Intermediate subclasses have been calculated by averaging between subsequent spectral types. The second dimension of the fitting matrix is created by progressively reddening all templates in the grid by steps of 0.1 visual magnitude, using the extinction law of \citet{Fitzpatrick1999}. 

An analogous grid was constructed that is composed of field (1-5~Gyr) template spectra of G and K stars \citep{Rayner2009} retrieved from the IRTF Spectral Library\footnote{Available at http://irtfweb.ifa.hawaii.edu/$\sim$spex/IRTF\_Spectral\_Library/}, together with M and L dwarf optical standards \citep{Burgasser2004,Burgasser2006,Burgasser2008} retrieved from the SpeX Prism Spectral Libraries\footnote{Available at http://www.browndwarfs.org/spexprism/}, all taken with the SpeX spectrograph \citep{Rayner2003} mounted on the 3~m NASA Infrared Telescope Facility. These are reddened in the same way as described for the grid of young objects, and the grid is smoothed by calculating intermediate spectral types from the averages of the templates. This secondary grid is used to double-check \emph{\`a posteriori} that the youth indicators described above are present in the candidate spectra. 

   \begin{table*}
   \centering             
\caption{Empirical grid of young template spectra.}            
\begin{tabular}{l c c c c c c c c}        
\hline \hline
2MASS Designation & Region & Optical SpT & Ref.\tablefootmark{a} & A$_{\emph{V}}$\tablefootmark{b} (mag) &  (\emph{J}$-$\emph{H})$_{2MASS}$\tablefootmark{c}  & (\emph{H}$-$\emph{K$_{s}$})$_{2MASS}$\tablefootmark{c} &  (\emph{J}$-$\emph{H})$_{WIRCam}$\tablefootmark{d}  & (\emph{H}$-$\emph{K$_{s}$})$_{WIRCam}$\tablefootmark{d}  \\
\hline
J03442812+3216002  	&	IC348  & M3.25   & 1	  & 2.0    &  0.69   &    0.22    &  0.70  &  0.20	   \\
J04311578+1820072   	&	Taurus & M4.25   & 2	  & 0.4    &  0.63   &    0.26    &  0.63  &  0.24	   \\
J04183030+2743208 	&	Taurus & M5.5	 & 2	  & 0.6    &  0.56   &    0.29    &  0.54  &  0.27	   \\
J04180796+2826036       &	Taurus & M6	 & 3,4,5  & 1.6    &  0.55   &    0.32    &  0.50  &  0.31	   \\
J04394748+2601407 	&	Taurus & M7	 & 6,2    & 5.4    &  0.60   &    0.38    &  0.55  &  0.37	   \\
J04305718+2556394 	&	Taurus & M8.25   & 2	  & 0.7    &  0.66   &    0.46    &  0.60  &  0.45	   \\
J04190126+2802487 	&	Taurus & M9	 & 7	  & 1.3    &  0.74   &    0.56    &  0.65  &  0.55	   \\
J01415823$-$4633574    &	field  & L0	 & 8	  & 0.0    &  1.01   &    0.65    &  0.91  &  0.65	   \\
J22081363+2921215     	&	field  & L3	 & 8	  & 0.0    &  1.19   &    0.78    &  1.04  &  0.78	    \\
J05012406$-$0010452    &	field  & L4	 & 8	  & 0.0    &  1.31   &    0.82    &  1.12  &  0.82	    \\
J03552337+1133437     	&	field  & L5	 & 8	  & 0.0    &  1.57   &    1.07    &  1.42  &  1.08	    \\
\hline
 \label{temp} 
\end{tabular}
\tablefoottext{a}{Spectral types from: 1: \citet{Luhman2003ic348}, 2: \citet{ Briceno2002}, 3:\citet{Strom1994}, 4:\citet{Luhman1998}, 5:\citet{Briceno1998}, 6:\citet{Martin2001}, 7: \citet{Luhman2003taurus}, 8: \citet{Cruz2009}. }
\tablefoottext{b}{Visual extinction determined from the \emph{J}$-$\emph{H}~vs.~\emph{H}$-$\emph{K$_{s}$} diagram, except for the field L dwarfs.}
\tablefoottext{c}{Colours computed from the dereddened spectra in the 2MASS photometric system.}
\tablefoottext{d}{Colours computed from the dereddened spectra in the WIRCam/CFHT Vega system.}
\end{table*}

\subsubsection{Fitting technique}
For each candidate spectrum, we determine the best fit in spectral type and extinction by minimizing a goodness-of-fit statistic between the candidate and each template \citep{Cushing2008}, defined as
   \begin{equation}  
GF = \sum_{\lambda} w_{\lambda} \left(\frac{f^{candidate}_{\lambda} - F^{template}_{\lambda}}{\sigma^{candidate}_{\lambda}}\right)^{2}
   \end{equation}
where $w_{\lambda}$ is the weight of each spectral element \citep[taken as its spectral width as in][]{Cushing2008}, $f_{\lambda}$ and $F_{\lambda}$ are the flux densities of the candidate and template, respectively, and $\sigma_{\lambda}$ is the noise for the candidate spectrum. Regions of the spectra dominated by telluric absorption were excluded from the fit, which was done over the wavelength ranges 1.45$-$1.8 and 2$-$2.45~$\mu$m. Since the ISAAC spectra cover only the \emph{H} and \emph{K} bands, we have used the photometric point of the \emph{J} band in the fit, which is compared to the spectro-photometric magnitude derived by each template in the grid, and given a weight that is the sum of the individual weight values in the \emph{J} band of the template. This approach is necessary in order to minimise the effect the degeneracy between extinction and spectral type have on the overall slope of a young late-M and L-type spectrum \citep[e.g.,][]{Allers2007}. 

\subsubsection{Testing the fitting method}
\begin{figure}
   \centering
\includegraphics[width=\columnwidth]{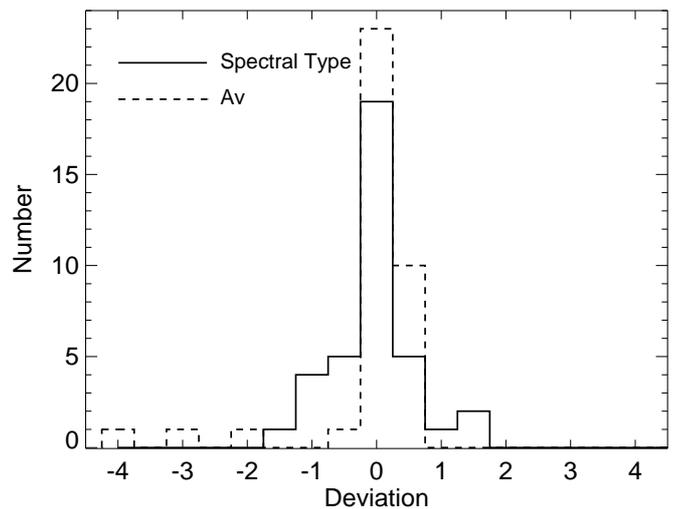}
\caption{Histogram of the deviation between the near-IR spectral type (solid line) and the extinction (dotted line) derived with our fitting technique for young members of Taurus and IC~348, and the spectral type derived from the optical and the extinction derived from the near-IR colour-colour magnitude diagram.}
\label{hist1}
\end{figure}

We tested this technique by determining the near-IR spectral type and extinction for the 30 aforementioned confirmed young members of Taurus and IC~348, where the \emph{J}-band part of each spectrum is replaced by its integrated flux to mimic our ISAAC dataset. The vast majority of the derived spectral types agree to better than 1 subclass with those determined from the optical, and to better than 1 magnitude of A$_{\emph{V}}$ from the extinction values determined from the near-IR colour-colour diagram, as shown in the histogram of the parameter's deviation in Fig.~\ref{hist1}. We take these to be the intrinsic errors in the fitting method and to confirm that an accurate spectral type can be derived when replacing the information of the \emph{J} band spectra with its integrated flux. Therefore, the final uncertainties in the spectral type and extinction derived for the $\rho$~Oph candidate members observed with ISAAC/VLT are taken to be 1 spectral subclass and 1~mag. 

\begin{table*}
\centering             
\caption{Brown dwarfs in the literature.}            
\begin{tabular}{l l c c c c c}        
\hline \hline
2MASS Designation & Region & Opt SpT\tablefootmark{a} & Ref.\tablefootmark{a} &   Other name & NIR SpT\tablefootmark{b} & A$_{\emph{V}}$\tablefootmark{b} (mag) \\
\hline
J11082927$-$7739198   	&	Chamaeleon I  & M7.25      & 1		& [NC98]~Cha~HA~11	&  M5.5   	&	1.5	\\
J11102226$-$7625138   	&	Chamaeleon I  & M8           & 1		& CHSM~17173    		&  M7.5   	&	0.8	\\
J11395113$-$3159214 	&	Twa Hydra	& M8           & 2   	& TWA~26  		 	&  M7.25  	&	1.8	\\
J11100934$-$7632178    &	Chamaeleon I  & M9.5	   & 3   	& OTS44    			&  M9.25  	&	2.6	\\
~				 	&	Chamaeleon I	& L0   	   & 4  	& [L2007b]~Cha J11070768$-$7626326 &  M9  	 &	0.8	\\
\hline
\label{knownBD} 
\end{tabular}

\tablefoottext{a}{Spectral types from: 1: \citet{Luhman2004cha}, 2: \citet{Rice2010} , 3: \citet{Luhman2004ots44} , 4: \citet{Luhman2008}. }
\tablefoottext{b}{Near-IR spectral type and extinction derived from numerical fitting.}
\end{table*}

We repeated the test for the known brown dwarfs in the literature observed during our ISAAC run.  Four out of the five brown dwarfs we observed have a spectral type solution from our fitting technique that agrees to within one subclass in respect to their published optical spectral type. For the brown dwarf Cha~H$\alpha$~11 we find a spectral type that is $\sim$2 subclasses earlier. The derived values of extinction show a dispersion of $\lesssim$2.6~mag to those published, which are zero for all targets. This could be explained by our reliance on 2MASS photometry for the relative calibration of the spectral bands, propagating the photometric errors to the spectrum. Furthermore, particular geometries of disks around the brown dwarfs could cause variability at near-IR wavelengths, as well as contribute to the \emph{K}-band overall flux, which would impact the extinction derived, but not significantly affect the overall spectral features. Table~\ref{knownBD} summarises the properties of these objects as retrieved from the literature, as well as the parameters derived from our spectral fitting method. 

\subsection{Spectral classification in the optical}

The most distinct feature in the optical spectra of M dwarfs are the absorption bands from titanium oxide (TiO) and vanadium oxide (VO), which are very sensitive to temperature, increasing in strength as we move to later spectral types in a smooth progression \citep[e.g., ][]{Kirkpatrick1993}. 
To determine the spectral type and extinction of our targets, we employ the same algorithm as is described in the previous section, using the goodness-of-fit statistic to determine the best match between the candidate member and the grid of templates. The grid is composed of young objects that are members of Chamaeleon  \citep[$\sim$1Myr, ][]{Luhman2004cha}, which have been shown to be suitable for relative spectral classification of young M dwarfs in star-forming regions \citep[see Table~2 in ][]{Riddick2007}. The templates are reddened progressively in steps of 0.1 visual magnitude, using the extinction law of \citet{Fitzpatrick1999}. The typical errors of this method in the spectral type domain we are studying (late M to early L) are of the order of 1 subclass in spectral type and 1 magnitude of extinction \citep[e.g.,][]{Guieu2006,Riddick2007}.

%__________________________________________________________________

\section{Results} \label{results}
\begin{figure}
   \centering
 \includegraphics[width=\linewidth]{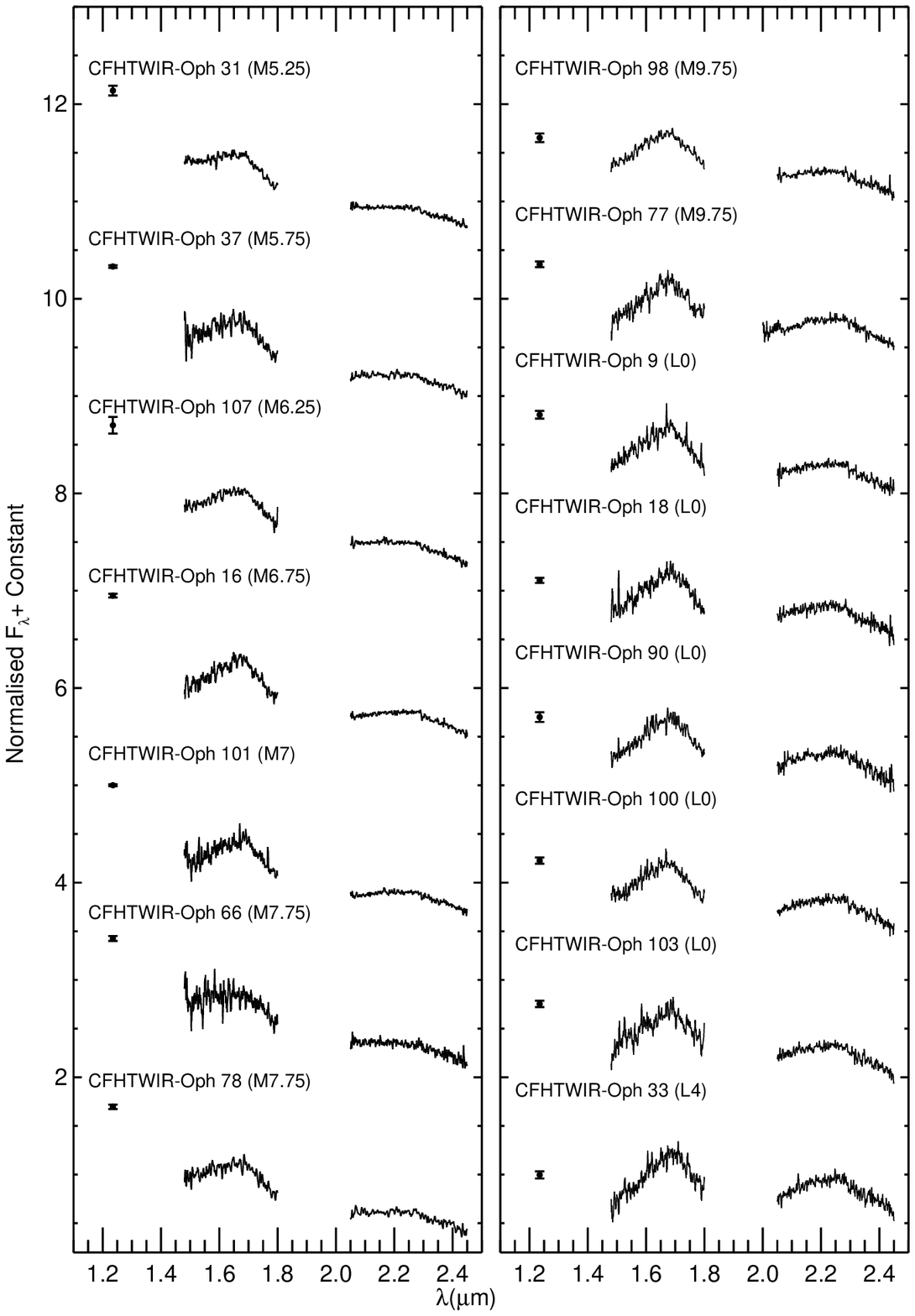}
   \caption{ISAAC/VLT low-resolution spectra of the new confirmed members in $\rho$~Oph. All spectra are corrected for extinction with the values found through the numerical fitting procedure.}
   \label{allspecir}
    \end{figure}

\begin{figure}
   \centering
 \includegraphics[width=\linewidth]{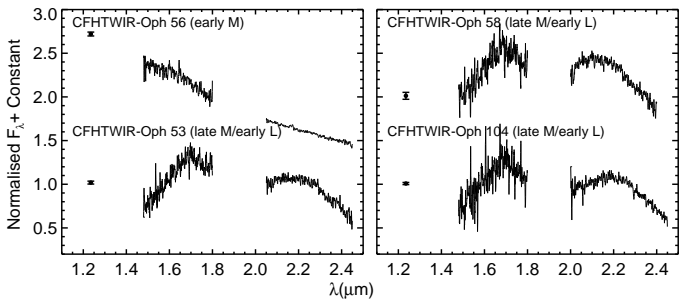}
   \caption{ISAAC/VLT low-resolution spectra of the members with uncertain spectral type.}
   \label{uncertainspt}
    \end{figure}

\begin{figure}
   \centering
\includegraphics[width=\columnwidth]{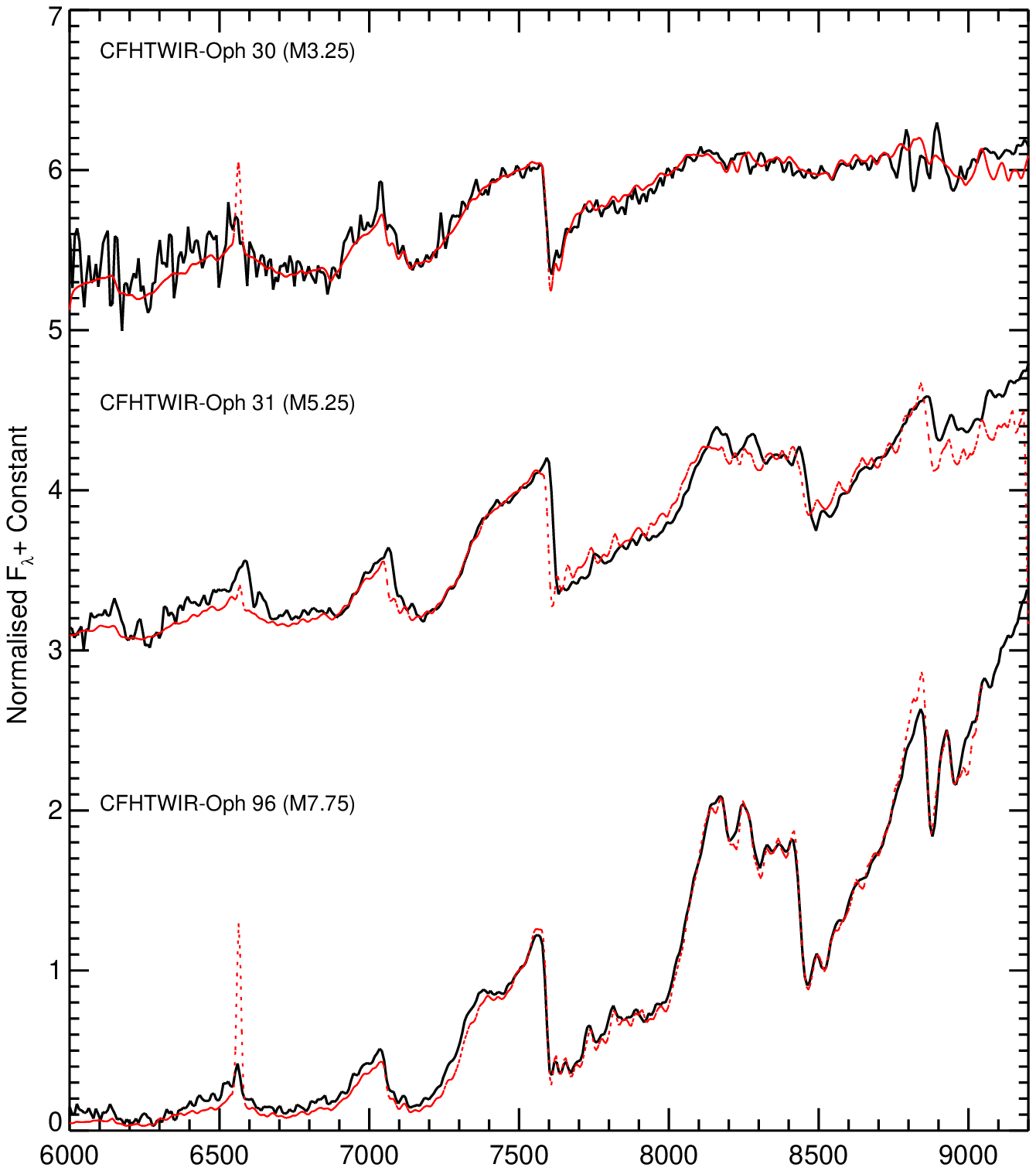}
\caption{Osiris/GTC  optical spectra of three confirmed members in $\rho$~Oph. All spectra are corrected for extinction with the values found through the numerical fitting procedure. The best-fitting template is overplotted in red.}
\label{allspecop}
\end{figure}

The numerical spectral fitting technique has been applied to all candidate members observed with ISAAC/VLT and Osiris/GTC, and the solution for spectral type and extinction is summarised in Table~\ref{result1}. The final dereddened spectra are shown in Figs.~\ref{allspecir}, \ref{uncertainspt}, and \ref{allspecop}. In the following sections we discuss the fits to the candidates in more detail, including the membership assessment and the level of contamination in the observed sample.
   \begin{table*}   
   \centering             
   \caption{Spectral type and A$_{\emph{V}}$ determined through numerical spectral fitting.}           
   \centering             
      \begin{tabular}{l l l l l l c c c c}       
   \hline      \hline
CFHTWIR-Oph\tablefootmark{a} & RA & Dec & $\textit{J}$ (mag) & $\textit{H}$ (mag) &  $\textit{K$_{s}$}$ (mag) &  SpT\tablefootmark{b}  &  A$_{\emph{V}}$ (mag)\tablefootmark{b} & T (K)\tablefootmark{c} & Properties \\
   \hline                        
 \multicolumn{10}{c}{Confirmed new members} \\
  \hline  
9   	& 16:26:03.28 & $-$24:30:25.9 & 17.76 & 16.33 & 15.32 & L0 							& 6.6 		& 2250 	& \emph{H${_2}$O}, \emph{CO}  			\\ 
16  	& 16:26:18.58 & $-$24:29:51.6 & 17.09 & 14.92 & 13.55 & M6.75 						& 18.8		& 2908	& \emph{ex}, \emph{H${_2}$O}, \emph{var}	\\
18  	& 16:26:19.41 & $-$24:27:43.9 & 18.92 & 17.16 & 16.04 & L0 							& 7.8 		& 2250 	& \emph{ex}, \emph{H${_2}$O}, \emph{CO}  	\\ 
30   	& 16:26:36.82 & $-$24:19:00.3 & 16.24 & 14.35 & 13.06 & M3.25 						& 4.9      		& 3379	& \emph{ex}, \emph{var}, X-ray			\\
31   	& 16:26:37.81 & $-$24:39:03.3 & 14.65 & 13.45 & 12.66 & M5.25,M5.5					& 5.7, 8.9		& 3091	& \emph{ex}, \emph{VO}					\\
33   	& 16:26:39.69 & $-$24:22:06.2 & 18.16 & 16.74 & 15.68 & L4   						& 3.6 		& 1650	& \emph{H${_2}$O}, \emph{CO}   			\\
37  	& 16:26:40.84 & $-$24:30:51.1 & 17.32 & 14.77 & 13.18 & M5.75 (M7.5)\tablefootmark{d}	& 23.4 (22.)	& 3024 	&  \emph{ex}, \emph{H${_2}$O} 			\\
66  	& 16:27:14.34 & $-$24:31:32.0 & 18.42 & 16.53 & 15.30 & M7.75 (M4.5)\tablefootmark{d}	& 15.1 (17.6)	& 2753	& \emph{ex}, \emph{H${_2}$O}, \emph{var}	\\
77    & 16:27:25.64 & $-$24:37:28.6 & 18.23 & 16.54 & 15.36 & M9.75 (M7)\tablefootmark{d}   	& 10.0 (15.4) 	& 2263	& \emph{H${_2}$O}, \emph{CO}  			\\
78  	& 16:27:26.23 & $-$24:19:23.1 & 16.35 & 14.36 & 13.04 & M7.75						& 16.4		& 2753	& \emph{ex}, \emph{H${_2}$O}, \emph{var}	\\
90    & 16:27:36.59 & $-$24:51:36.1 & 16.83 & 15.65 & 14.85 & L0							& 2.4		& 2250	& \emph{ex}, \emph{H${_2}$O}, \emph{CO}  	\\
96   	& 16:27:40.84 & $-$24:29:00.8 & 14.60 & 13.76 & 13.19 & M7.75  						& 2.4    		& 2753	& \emph{ex}, \emph{VO}					\\
98   	& 16:27:44.20 & $-$23:58:52.4 & 17.08 & 15.78 & 14.98 & M9.75						& 3.0		& 2263	& \emph{ex}, \emph{H${_2}$O}, \emph{CO}  	\\
100 	& 16:27:46.54 & $-$24:05:59.2 & 17.94 & 16.34 & 15.26 & L0 							& 8.0		& 2250	&\emph{H${_2}$O}, \emph{CO}  			\\	
101 	& 16:27:47.25 & $-$24:46:45.9 & 19.26 & 16.48 & 14.78 & M7							& 25.8		& 2880	& \emph{H${_2}$O}, \emph{em}			\\
103  & 16:28:10.46 & $-$24:24:20.4 & 17.74 & 16.15 & 15.07 & L0							& 7.5		& 2250	& \emph{ex}, \emph{H${_2}$O}, \emph{CO}	\\
107  & 16:28:48.71 & $-$24:26:31.8 & 14.31 & 13.60 & 13.13 & M6.25						& 2.3		& 2963	& \emph{ex}							\\
   \hline                        
 \multicolumn{10}{c}{Members with uncertain spectral type} \\
  \hline  
53  	& 16:26:56.36 & $-$24:41:20.6 &  18.60 & 16.34 & 14.92 & L3?						& 8.6		&   	  	& \emph{ex}, \emph{H${_2}$O}, \emph{CO}	\\
56  	& 16:27:03.59 & $-$24:20:05.6 &  17.13 & 15.01 & 13.72 & M3?						& 20.1		&  		& \emph{ex}, \emph{var},	\emph{em}		\\
58  	& 16:27:05.94 & $-$24:18:40.3 &  17.05 & 15.77 & 14.85 & L3?						& 1.5		& 		& \emph{ex}, \emph{H${_2}$O}			\\
104  & 16:28:11.61 & $-$24:37:29.9 &  19.51 & 16.95 & 15.41 & L3?						& 10.5		& 		& \emph{ex}, \emph{H${_2}$O} 			\\ 
   \hline                        
 \multicolumn{10}{c}{Likely contaminants} \\
  \hline  
1   & 16:25:19.15 & $-$24:19:27.4 & 19.07 & 17.33 & 15.86 	&  ~ 		& ~  		& ~ 	& \emph{ex} 	\\
3   & 16:25:23.78 & $-$24:22:58.4 & 16.30 & 14.85 & 13.97 	&  M5 	& 9.7  	& ~ 	& \emph{Na~I} \\
6   & 16:25:46.63 & $-$24:23:36.4 & 18.87 & 16.69 & 15.12    &  ~   	& ~  		& ~ 	& \emph{ex} 	\\
7   & 16:25:46.98 & $-$24:11:54.1 & 18.34 & 16.85 & 15.68  	&  ~   	& ~  		& ~ 	& \emph{ex} 	\\
12 & 16:26:09.99 & $-$24:14:40.1 & 18.15 & 16.35 & 15.25 	&   M4  	& 15.3  	& ~ 	& \emph{Na~I} \\
15 & 16:26:16.33 & $-$24:39:30.7 & 15.79 & 14.15 & 13.11 	&   M6.5	& 12.4	& ~	& \emph{Na~I} \\ 
17 & 16:26:19.21 & $-$24:41:30.1 & 17.44 & 15.80 & 14.61 	&  ~ 		& ~  		& ~ 	& ~  			\\
19 & 16:26:22.27 & $-$24:37:09.5 & 19.15 & 17.10 & 15.86 	&  ~ 		& ~  		& ~ 	& ~  			\\
20 & 16:26:22.89 & $-$23:58:34.2 & 18.01 & 16.73 & 15.80	&  ~ 		& ~  		& ~ 	& ~  			\\
22 & 16:26:25.10 & $-$24:41:32.7 & 17.03 & 15.52 & 14.54	&  ~ 		& ~  		& ~ 	& ~  			\\
28 & 16:26:35.28 & $-$24:42:39.2 & 19.68 & 17.79 & 16.36 	&  ~ 		& ~  		& ~ 	& \emph{ex}  	\\
42 & 16:26:44.51 & $-$24:04:08.8 & 18.64 & 17.19 & 16.08 	&   ~ 	& ~  		& ~ 	& ~  			\\
52 & 16:26:55.54 & $-$23:57:36.5 & 16.02 & 15.08 & 14.39 	&   ~ 	& ~  		& ~ 	& \emph{ex}  	\\
69 & 16:27:18.53 & $-$24:07:22.0 & 14.76 & 13.55 & 12.79    &  M7.5	& 6.7	& ~	& \emph{Na~I}	\\ 
 \hline                                 
   \label{result1} 
   \end{tabular}
 \tablefoottext{a}{Name as in \citet{AlvesdeOliveira2010}, where additional information regarding each target is included.} 
\tablefoottext{b}{Based on the tests performed to the spectral fitting method, uncertainties on the spectral type and A$_{V}$ are estimated as 1 spectral subclass and 1~mag, respectively.} 
\tablefoottext{c}{See Sect.~\ref{secthr} where the temperature scale applied is explained.} 
\tablefoottext{d}{See Sect.~\ref{com} for the discussion.}
   \end{table*}

\subsection{New spectroscopically confirmed members of $\rho$~Oph}
By applying the numerical spectral fitting technique to all spectra, we classify 20 of the observed targets as new young members of the $\rho$~Ophiuchi cluster, with spectral types between M3 and L4. An additional target, \object{CFHTWIR-Oph~96} had been previously classified as a member through near-IR spectroscopy \citep[M8.25, A$_{V}$=1.10~mag,][]{AlvesdeOliveira2010}. In this work, we derive a more accurate optical spectral type (M7.75, A$_{V}$=2.4~mag), which agrees, within the errors, with the near-IR spectral type. We classify 14 targets as likely contaminants.

\subsubsection{Membership}
The youth of each target, hence membership, was assigned based on the presence of signatures of low-gravity or/and accretion disks, as well as each target's spatial location in relation to the cloud environment. For that end, the results from the spectroscopic follow-up were complemented with the available information in the literature, most of which had already been described and summarised in Table~1 of the WIRCam survey paper of $\rho$~Oph \citep{AlvesdeOliveira2010}. For targets with a near-IR spectral type later than $\sim$M3, the shape of the water-vapour absorption bands has been matched to that of spectra of young stellar objects with low gravity, providing a clear indication of their youth (denoted as \emph{H$_{2}$O} in Table~\ref{result1}). The detection of mid-IR excess in the \emph{Spitzer} photometry consistent with that expected for young objects surrounded by disks (\emph{ex} in Table~\ref{result1}) is another indication of membership. For these objects, accretion might still be on-going, so any signatures of Br~$\gamma$ emission are also indicative of youth, as well as H$_{2}$ emission arising from outflows. Furthermore, all objects selected as members do not show any sign of Na~I absorption, which is consistent with them having low gravity. Finally, the spatial location of an object also contains relevant information of whether it belongs to the cloud or not. For targets that do not show strong IR excess, the extinction derived in the numerical spectral fitting is not contaminated by the contribution of a disk, and should reflect the position of the target in respect to the depth of the cloud.  We compared that value to the extinction map of the Ophiuchus cloud provided by the COMPLETE\footnote{Available at http://www.cfa.harvard.edu/COMPLETE} project \citep{Ridge2006}, computed with the NICER algorithm \citep{Lombardi2001} using data from the \emph{Spitzer} Legacy Program "From Molecular Cores to Planet Forming Disks" \citep{Evans2003}. The extinction provided by the map is determined over a beam that is larger than our target observations and also over the entire depth of the cloud. Nonetheless, it still provides a good estimate of the environment for each candidate. All of our candidate members (except CFHTWIR-Oph~101, see discussion below) have extinctions that place them inside the cloud, when compared to the \emph{Spitzer} extinction map. That is not the case, for example, for the M dwarfs we listed in Table~\ref{result1} as likely contaminants, which have extinctions equal to or even greater than the expected values for the respective location placing them behind the cloud.

From the targets classified as members, sixteen (CFHTWIR-Oph~16, 18, 30, 31, 37, 53, 56, 58, 66, 78, 90, 96, 98, 103, 104, 107) show mid-IR excess indicative of a disk, as well as spectral features consistent with low gravity, which we consider the most constraining clues to their membership. In addition, five from those (CFHTWIR-Oph~16, 30, 56, 66, 78) have the near-IR variability typical of young stellar objects \citep{AlvesdeOliveira2008}. The membership of the five targets without mid-IR excess (CFHTWIR-Oph~9, 33, 77, 100, 101) is assigned based on complementary diagnostics and discussed below in more detail. 

\paragraph{CFHTWIR-Oph~9, 33, 77:} Candidate 9 is located in a region of the cloud with high extinction (A$_{V}$$\approx$17~mag), and the derived extinction value for this object is A$_{V}$$\simeq$6.6~mag, clearly placing it in the middle of the cloud. Its \emph{K} band spectra also clearly shows the CO overtones characteristic in spectra of early L dwarfs, which is consistent with its near-IR and optical photometric colours. CFHTWIR-Oph~33 is located close to the core of the cloud in a region with an A$_{V}$$\approx$24~mag, this target shows very peaked \emph{H} and \emph{K} band spectra characteristic of young objects, and with an A$_{V}$$\simeq$3.6~mag it is definitely not a background source. It is too bright to be a late L foreground field dwarf that its spectra indicate, further confirming its membership to the cluster. Similar to these two targets, CFHTWIR-Oph~77 is located in an area of high extinction (A$_{V}$$\approx$25~mag), and following the same argumentation, both the derived A$_{V}$ and spectral type are consistent with it being a member of the cloud. It could not be reproduced by any of the templates in the grid of field dwarfs. The spectral classification of this target is discussed further in Sect.~\ref{com}.  

\paragraph{CFHTWIR-Oph~100 and 101:} The target 100 is in a region of relatively low extinction (A$_{V}$$\simeq$5-8~mag), which places the object at the further edge of the cloud, when compared to the extinction derived by numerical fitting (A$_{V}$$\approx$8~mag). When derredening the colours of the target by this amount, the object still lies confidently above the main-sequence and is rather consistent with the expected colours of an early L young brown dwarf. The signal-to-noise ratio in the \emph{K} band is not enough to make a definite assessment of the Na~I feature. The CO overtone absorptions in the \emph{K} band further support its late spectral type, which again would be too bright to be a field dwarf behind the cloud. CFHTWIR-Oph~101 is the only target we classified as a member for which the derived extinction value exceeds that expected for the region from the extinction map. When looking at the near-IR images, it is, however, clear that differential extinction is present, since this part of the cloud shows a ridge-like structure. The target sits on one of the dense ridges, which seems to explain the high extinction, and has most likely been smoothed in the low spatial resolution of the extinction map. There is no indication of Na~I absorption, further supporting its low-gravity, while there is an indication of Br~$\gamma$ emission (2.16~$\mu$m) showing that accretion may be on-going. Finally, the peaked shape of the \emph{H} band is not reproduced by any spectra of the grid of field dwarfs. Based on these arguments, we classify both targets as members of the $\rho$~Ophiuchi cluster. 

\subsubsection{Comments on individual sources}\label{com}
\paragraph{CFHTWIR-Oph~30:} \label{var}
The candidate member CFHTWIR-Oph~30 has been previously associated with the cloud by several diagnostics, including mid-IR excess emission consistent with that of YSOs \citep[e.g.,][]{Bontemps2001,Gutermuth2009} and X-ray emission \citep{Gagne2004}. More recently, \citet{AlvesdeOliveira2008} found this object to show extreme variability in the near-IR, with a brightening of 3.5 magnitudes in the \emph{K} band in only one year, which the authors suggest could indicate an EXor-type object, a class of young variable stars where the large magnitude variations are thought to be caused by massive infall of circumstellar material onto the central star \citep[e.g.,][]{Herbig2008}. Based on the acquisition images of the ISAAC dataset, we have estimated a magnitude of \emph{K}$=$14.54$\pm$0.3, which would indicate that in 2010 the target was again approaching its minimum state. As a result, the necessary spectroscopic integration times have been underestimated and the signal-to-noise of the ISAAC spectra is too low for a reliable spectral fitting. This is underpinned further by the large uncertainty of the near-IR magnitudes at the time of the spectroscopic observations, crucial for calibrating the relative spectral fluxes of the various bands. This target was also observed spectroscopically in the near-IR by \citet{Geers2011} (source \#3 in their Table~1) and classified as non-substellar. We obtained an optical spectrum of CFHTWIR-Oph~30 with GTC, allowing more reliable analysis. We derive a spectral type of M3.5 with a  visual extinction of 4.7 magnitudes (see Fig.\ref{allspecop}). There is no indication of H$\alpha$ emission in the optical spectrum, which could support the scenario where the YSO is at its rest accretion state. 

\begin{figure}
   \centering
\includegraphics[width=\columnwidth]{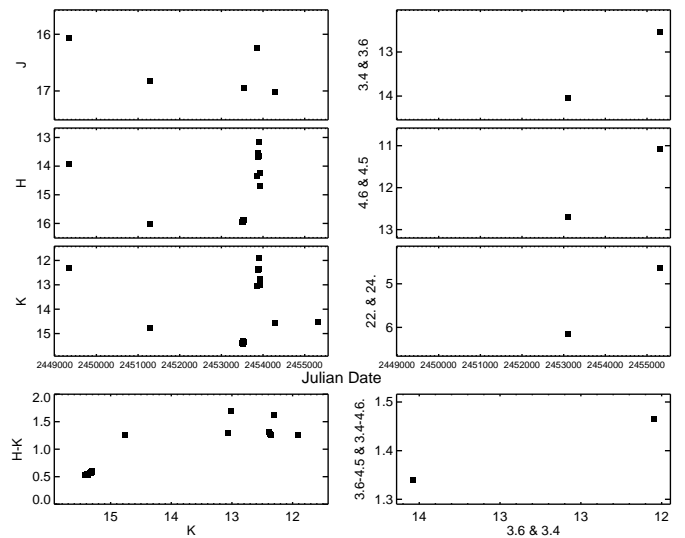}
\caption{Light curve for CFHTWIR-Oph~30 combining all the IR measurements available (SQIID/KPNO, 2MASS, WFCAM/UKIRT, WIRCam/CFHT, MOIRCS/Subaru, IRAC/Spitzer, and WISE), where uncertainties are typically $\lesssim$0.1~mag. The magnitudes have not been corrected to a unique photometric system, though this cannot account for the large variations that can only be explained by intrinsic variability.}
\label{oph30}
\end{figure}

We have compiled the available IR measurements from the literature, which are described in the following articles: \citet[][SQIID/KPNO]{Barsony1997}, \citet[][2MASS]{Cutri2003},  \citet[][WFCAM/UKIRT]{AlvesdeOliveira2008}, \citet[][WIRCam/CFHT]{AlvesdeOliveira2010}, \citet[][MOIRCS/Subaru]{Geers2011}, \citet[][IRAC/Spitzer]{Gutermuth2009}, and \citet[][WISE]{Wright2010}. The resulting light curve for the different bands is shown in Fig.~\ref{oph30}. Even though we did not correct the magnitudes to account for the differences between photometric systems, this cannot account for the large amplitude that is seen, and it can only be explained by intrinsic variability. This plot strengthens the rare colour effect already measured by \citet{AlvesdeOliveira2008}, with the star bluer when fainter and the cause of variation possibly related to the star's disk. For example, as a clump of material or warp in the disk moves out of the observers line-of-sight, the star-disk system becomes fainter in the IR since less light is being absorbed and reprocessed by the circumstellar disk. Disks radiate strongly at longer wavelengths, so when the warped inner disk region rotates out of the observed fraction of the star, the stellar colours become bluer, approaching its intrinsic colours. However, the intrinsic fluxes of an M3.5 YSO are much brighter than what is observed at its minimum. Indeed, its optical spectral type would indicate that the nominal state of the target is at its brightest stage. If taking this into account, the variability could instead be explained by a companion object, since the percentage of coverage of a cold spot on the surface needed to reproduce such behaviour is not physical when considering the magnitude variations. Then, the observed behaviour could be due to a system consisting of two objects with similar spectral type (i.e. similar intrinsic luminosity), one having strong IR excess and undergoing variable circumstellar extinction by the disk (i.e., close to edge-on), the other component either disk free or at least not obscured by its disk (e.g. different inclination). Then, the whole system would become fainter in the near-IR when the disk obscures the primary. At that time the secondary would start to dominate the near-IR integrated light. If the secondary has no (or a smaller) IR excess, its near-IR colours will be bluer than that of the primary, and the system would in turn become bluer when fainter in the near-IR. A detailed modelling of the system is beyond the scope of this paper, and will most likely be addressed by dedicated monitoring surveys such as YSOVAR\footnote{Available at http://ysovar.ipac.caltech.edu.} \citep{Morales2011}.

\paragraph{CFHTWIR-Oph~37, 66, 77:} For all the targets we analysed, the near-IR numerical fit is done both with the \emph{J} band photometric point and without it, for which we obtain the same solution, confirming that the spectral shape of the \emph{H} and \emph{K} bands is the dominant component of the fit, as opposed to the \emph{JHK} colours on which the spectra are calibrated. This is, however, not the case for three targets (CFHTWIR-Oph~37, 66, 77) where we obtained two different solutions depending on whether or not we included the comparison between the \emph{J} photometric point of the target to the integrated flux of the template's \emph{J} spectra in the fit. The different solutions are in fact correlated, since they are extreme cases of the degeneracy between extinction and spectral type, which exists at this low resolution \citep[e.g.,][]{Allers2007}, and also at a modest signal-to-noise of the targets. In the case of CFHTWIR-Oph~37 and 66, their spectral energy distribution (SED) shows a strong IR excess up to 24~$\mu$m, meaning the contribution from the disk emission could affect the determination of the extinction. If we investigate their near-IR colours, CFHTWIR-Oph~37 is consistent with the earlier spectral type solution (M5.75, A$_{V}$=23.4~mag), while the faint magnitudes of CFHTWIR-Oph~66 are in agree with the later spectral type solution (M7.75, A$_{V}$=15.1~mag). We adopt these classifications throughout the rest of the paper. CFHTWIR-Oph77 does not show mid-IR excess, and its colours are consistent with a later spectral type, which is also supported by the appearance of the CO overtone absorption in its \emph{K} band spectra, and therefore we adopt the spectral type solution M9.75 with A$_{V}$=10.0~mag. 

\paragraph{CFHTWIR-Oph~53, 56, 58, 104:} These targets are assigned a spectral type that carries a larger uncertainty than the other candidates, due to both the poor signal-to-noise ratio and strong mid-IR excess (see Fig.~\ref{uncertainspt}). CFHTWIR-Oph~53, 58, and 104 show strong water vapour absorption in the \emph{H} band, as well as pronounced CO overtone absorption in the \emph{K} band, and are largely consistent with a late M to early L spectral type. The signal-to-noise in the \emph{H} band is too low to assess the quality of the fit, and the solutions presented rely mainly on the \emph{K} band features. CFHTWIR-Oph~56 appears to have an earlier spectral type than M3, the hottest template in our grid. They all have clear signatures of youth and are certainly members of the cluster. We include the best spectral solution found in Table~\ref{result1}, but caution that the error in spectral type is significantly higher than for the other targets. We take a conservative approach and do not include them in the analysis of the brown dwarf population in Sects.~\ref{discussion1} and \ref{discussion2}. 

\subsection{Likely contaminants}
We classify fourteen candidates in our sample as contaminants, which are most likely background giants or field M dwarfs. All these targets (CFHTWIR-Oph~1, 3, 6, 7, 12, 15, 17, 19, 20, 22, 28, 42, 52, and 69) have a non-negligible amount of extinction, so they cannot be foreground stars. For most of the targets (CFHTWIR-Oph~1, 6, 7, 17, 19, 20, 22, 28, 42, 52), there is no sign of water absorption features in their spectra, implying a spectral type that is earlier than M3. However, a young object ($\sim$1Myr) at the distance of $\rho$~Oph ($\sim$130~pc), ought to be much brighter than the faint magnitudes these targets show. Therefore, they are most likely giants and AGB stars, also known to have mid-IR excess, which are seen through the cloud. The targets classified as field M dwarfs (CFHTWIR-Oph~3, 12, 15, 69), show spectral features that are better matched by our grid of template field dwarfs, such as a plateau in the shape of the \emph{H} band, and the Na~I absorption feature characteristic of older objects. Furthermore, they do not show any other sign of youth, such as mid-IR, and all have extinction solutions that place them behind the cloud when compared to those predicted by the extinction maps. 

\subsection{Effectiveness of the photometric selection}
We combined the results for all the candidate members selected in our WIRCam/CFHT survey that have been observed spectroscopically. These include seven confirmed members presented in our previous spectroscopic follow-up \citep{AlvesdeOliveira2010}, where we found four contaminants (these were excluded from the list of candidate members of the cluster). Furthermore, \citet{McClure2010} confirm the membership of six of our candidate members. In this study, we have confirmed as members of the cluster 20 additional targets. The candidates classified as contaminants total 18. In absolute terms, this implies a success rate of 65\%. If we take all the candidate members within our photometric completeness limits into account (see Fig.~\ref{cmd}), by including 17 targets that have not been observed spectroscopically, we derive lower and upper limits of the success rate of the WIRCam/CFHT survey of $\rho$~Oph of 49\% and 74\%, respectively. This agrees with previous estimates in surveys of other clusters \citep[e.g.,][]{Oliveira2009,Scholz2009,Lodieu2011b}, and reenforces both the need for spectroscopic confirmation of candidate members selected from photometry and the limitations of previous IMF studies based only on photometric selected samples.

%__________________________________________________________________

\section{The brown dwarf population in $\rho$~Ophiuchi} \label{discussion1}

\subsection{An updated census} \label{bdcensus}
Despite being one of the closest and youngest star-forming regions where brown dwarfs should be more easily detected, the high and variable extinction of the $\rho$~Ophiuchi cluster has meant that its substellar population remained largely unknown. We compiled a list of all brown dwarf members of $\rho$~Oph confirmed spectroscopically to date, which is an updated version of Table~4 in \citet{AlvesdeOliveira2010}. At the age of $\rho$~Oph ($\sim$1Myr), the evolutionary models of the Lyon group, combined with the empirical temperature to spectral type relation determined by \citet{Luhman2003ic348}, imply that the mass limit for hydrogen burning corresponds to a spectral type of $\sim$M6.25 \citep{Luhman2007}. Given the limited accuracy in near-IR low-resolution spectroscopy, typically of one subclass, members classified with a spectral type of M6 are listed in the table of brown dwarfs as members at the substellar limit. To be consistent with other studies of star-forming regions and to facilitate comparison of properties across different populations, we label young objects with spectral types later than M6 as brown dwarfs. By this criterion, at the time of the latest review of the cluster \citep{Wilking2008}, only 11 brown dwarf members were known in Ophiuchus. With the WIRCam survey and the subsequent spectroscopic follow-up \citep[and this work]{AlvesdeOliveira2010}, we have confirmed 19 new brown dwarfs (22 if we include CFHTWIR-Oph~53, 58, and 104, which have uncertain spectral types between late M and early L), a threefold increase in the substellar population of $\rho$~Oph. Based on the optical classification of \citet{Comeron2010}, we do not include GY~10 (M5.5) in the substellar census and update the spectral type of GY~11. Other recent contributions to the census of the cluster have been done by \citet{McClure2010}, where spectral types were derived up to M6 for several members, and the confirmation of one brown dwarf member by \citet{Geers2011}.\\

\begin{landscape}
   \begin{table}   
   \tiny
   \caption{Spectroscopically confirmed brown dwarfs in $\rho$~Oph.}           
   \centering             
       \begin{tabular}{l l l l l l c c c c c c c c}       
   \hline      \hline
Identifier & RA & Dec & SpT  &  A$_{\emph{V}}$ (mag) & References\tablefootmark{a} & $\textit{J}$ (mag) & $\textit{H}$  (mag) &  $\textit{K$_{s} $}$ (mag) &  [3.6] (mag) &  [4.5] (mag) &  [5.6] (mag) &  [8.0] (mag) &  [24] (mag)  \\
   \hline          
  \hline                        
 \multicolumn{14}{c}{Members at the substellar limit} \\
  \hline  
\object{GY92~15}	       & 16:26:22.98 & -24:28:46.1 & M6    & 10.7 & 8	     & 14.93$\pm$0.04 & 12.80$\pm$0.02 & 11.53$\pm$0.02 & 10.41$\pm$0.06 & 10.00$\pm$0.06 &  9.50$\pm$0.07 &  8.46$\pm$0.09 &  6.11$\pm$0.63 \\   	 
\object{GY92~109}	       & 16:26:42.89 & -24:22:59.1 & M6    & 13.5 & 8	     & 15.33$\pm$0.05 & 12.82$\pm$0.03 & 11.44$\pm$0.02 & 10.18$\pm$0.06 &  9.63$\pm$0.06 &  9.11$\pm$0.06 &  8.35$\pm$0.09 &  6.74$\pm$0.72 \\  	 
\object{GY92~154}	       & 16:26:54.79 & -24:27:02.1 & M6    & 20.1 & 8	     & 17.92$\pm$0.05 & 14.91$\pm$0.05 & 12.87$\pm$0.05 & 11.13$\pm$0.06 & 10.51$\pm$0.06 & 10.00$\pm$0.06 &  9.34$\pm$0.09 &  5.85$\pm$0.63 \\   	 
\object{WL21}	       & 16:26:57.33 & -24:35:38.7 & M6    & 23.8 & 8	     & 19.00$\pm$0.05 & 15.07$\pm$0.05 & 12.82$\pm$0.05 & 11.41$\pm$0.10 & 10.41$\pm$0.06 &  9.96$\pm$0.07 &  9.44$\pm$0.09 &  6.93$\pm$0.65 \\    
\object{GY92~171}	       & 16:26:58.41 & -24:21:30.0 & M6    & 6.6  & 8	     & 16.01$\pm$0.07 & 13.11$\pm$0.02 & 11.46$\pm$0.02 & 10.09$\pm$0.06 &  9.55$\pm$0.06 &  9.11$\pm$0.06 &  8.50$\pm$0.09 &  5.80$\pm$0.63 \\ 	 
\object{GY92~204}	       & 16:27:06.60 & -24:41:48.8 & M6    & 0.5  & 5	     & 12.43$\pm$0.02 & 11.40$\pm$0.02 & 10.77$\pm$0.02 &  9.95$\pm$0.05 &  9.67$\pm$0.06 &  9.32$\pm$0.07 &  8.65$\pm$0.09 &  5.23$\pm$0.63 \\   	
\object{ISO-Oph~160}        & 16:27:37.42 & -24:17:54.9 & M6    & 6.0  & 5	     & 14.15$\pm$0.03 & 12.76$\pm$0.03 & 11.95$\pm$0.03 & 11.00$\pm$0.05 & 10.61$\pm$0.06 & 10.13$\pm$0.06 &  9.35$\pm$0.09 &  6.74$\pm$0.63 \\   	 
\object{GY92~344}	       & 16:27:45.78 & -24:44:53.6 & M6    & 16.2 & 8	     & 17.74$\pm$0.05 & 14.80$\pm$0.05 & 12.61$\pm$0.05 &  9.70$\pm$0.06 &  8.71$\pm$0.06 &  7.97$\pm$0.06 &  7.27$\pm$0.09 &  3.88$\pm$0.63 \\   	 
\object{GY92~350}	       & 16:27:46.29 & -24:31:41.2 & M6    & 7.0  & 5	     & 13.83$\pm$0.02 & 12.21$\pm$0.02 & 11.32$\pm$0.02 & 10.51$\pm$0.05 & 10.15$\pm$0.06 &  9.73$\pm$0.06 &  8.84$\pm$0.09 &  6.16$\pm$0.63 \\   	 
\object{GY92~371AB}        & 16:27:49.77 & -24:25:22.2 & M6    & 5.4  & 8	     & 12.78$\pm$0.02 & 11.12$\pm$0.02 & 10.16$\pm$0.02 &  9.18$\pm$0.06 &  8.76$\pm$0.06 &  8.39$\pm$0.07 &  7.84$\pm$0.09 &  5.86$\pm$0.63 \\   	 
\object{GY92~397}	       & 16:27:55.24 & -24:28:39.7 & M6    & 5.0  & 8	     & 13.04$\pm$0.02 & 11.59$\pm$0.02 & 10.79$\pm$0.02 &  9.98$\pm$0.05 &  9.62$\pm$0.06 &  9.17$\pm$0.06 &  8.35$\pm$0.09 &  5.69$\pm$0.63 \\   	 
\object{GY92~450}	       & 16:28:03.56 & -24:34:38.6 & M6    & 20.5 & 8	     & 19.17$\pm$0.05 & 15.19$\pm$0.05 & 13.11$\pm$0.05 & 11.72$\pm$0.05 & 11.41$\pm$0.06 & 11.12$\pm$0.07 & 11.05$\pm$0.09 &  ...           \\       
\object{ISO-Oph~193}       & 16:28:12.72 & -24:11:35.6 & M6    & 7.5  & 5	     & 13.61$\pm$0.02 & 12.02$\pm$0.02 & 11.09$\pm$0.02 & 10.18$\pm$0.06 &  9.91$\pm$0.06 &  9.43$\pm$0.07 &  8.63$\pm$0.09 &  5.79$\pm$0.63 \\     
  \hline                        
  \multicolumn{14}{c}{Brown dwarfs} \\
  \hline  
\object{CFHTWIR-Oph~107} & 16:28:48.71 & -24:26:31.8 & M6.25 & 2.3  & this work & 14.31$\pm$0.05 & 13.60$\pm$0.05 & 13.13$\pm$0.05 & 12.55$\pm$0.06 & 12.26$\pm$0.06 & 11.88$\pm$0.07 & 11.29$\pm$0.09 &  9.73$\pm$0.68 \\   
\object{CFHTWIR-Oph~4}   & 16:25:32.41 & -24:34:05.2 & M6.5  & 2.5  & 7	     & 14.88$\pm$0.05 & 13.93$\pm$0.05 & 13.34$\pm$0.05 & 12.71$\pm$0.06 & 12.53$\pm$0.06 & 12.37$\pm$0.11 & 12.31$\pm$0.46 &  ...           \\   	
\object{CFHTWIR-Oph~106} & 16:28:29.93 & -24:54:06.4 & M6.5  & 4.9  & 7	     & 15.36$\pm$0.05 & 14.51$\pm$0.05 & 13.85$\pm$0.05 & 13.05$\pm$0.06 & 12.57$\pm$0.06 & 12.11$\pm$0.07 & 11.47$\pm$0.10 &  ...           \\   	
\object{CRBR~2322.3-1143}         & 16:26:23.78 & -24:18:31.4 & M6.7  & 8.6  & 4	     & 16.28$\pm$0.05 & 14.62$\pm$0.05 & 13.57$\pm$0.05 & 12.53$\pm$0.06 & 12.25$\pm$0.06 & 12.02$\pm$0.08 & 11.49$\pm$0.10 &  8.17$\pm$0.63 \\   	  
\object{CFHTWIR-Oph~16}  & 16:26:18.58 & -24:29:51.6 & M6.75 & 18.8 & this work & 17.09$\pm$0.05 & 14.92$\pm$0.05 & 13.55$\pm$0.05 & 12.14$\pm$0.06 & 11.61$\pm$0.06 & 11.00$\pm$0.08 & 10.18$\pm$0.16 &  6.99$\pm$0.63 \\   	  
\object{GY92~202}	       & 16:27:05.98 & -24:28:36.3 & M7(M6.5)    & 13.0 & 2,3      & 16.79$\pm$0.05 & 14.58$\pm$0.05 & 13.15$\pm$0.05 & 11.78$\pm$0.06 & 11.35$\pm$0.06 & 10.84$\pm$0.08 & 10.19$\pm$0.09 &  7.08$\pm$0.63 \\   	  
\object{CFHTWIR-Oph101} & 16:27:47.25 & -24:46:45.9 & M7    & 25.8 & this work & 19.26$\pm$0.05 & 16.48$\pm$0.05 & 14.78$\pm$0.05 & 13.35$\pm$0.06 & 13.01$\pm$0.06 & 12.72$\pm$0.08 & 12.65$\pm$0.13 &  ...           \\  
\object{CFHTWIR-Oph57}  & 16:27:04.02 & -24:02:46.9 & M7.25 & 6.1  & 7	     & 15.06$\pm$0.05 & 13.92$\pm$0.05 & 13.18$\pm$0.05 & 12.53$\pm$0.07 & 12.37$\pm$0.07 & 12.14$\pm$0.08 & 11.96$\pm$0.11 &  ...           \\  
\object{CRBR~2317.3-1925}         & 16:26:18.82 & -24:26:10.5 & M7.5(M5.5,M7)  & 10.0 & 2,3,5    & 14.84$\pm$0.04 & 13.20$\pm$0.03 & 12.14$\pm$0.02 & 11.07$\pm$0.06 & 10.48$\pm$0.06 &  9.79$\pm$0.07 &  8.74$\pm$0.09 &  6.73$\pm$0.64 \\   	  
\object{CFHTWIR-Oph47}  & 16:26:52.26 & -24:01:46.8 & M7.5  & 5.6  & 7	     & 15.94$\pm$0.05 & 14.75$\pm$0.05 & 13.97$\pm$0.05 & 13.35$\pm$0.07 & 13.17$\pm$0.07 & 13.05$\pm$0.11 & 12.96$\pm$0.22 &  ...           \\   	
\object{CFHTWIR-Oph66}  & 16:27:14.34 & -24:31:32.0 & M7.75 & 15.1 & this work & 18.42$\pm$0.05 & 16.53$\pm$0.05 & 15.30$\pm$0.05 & 14.07$\pm$0.06 & 13.49$\pm$0.06 & 13.06$\pm$0.10 & 11.90$\pm$0.11 &  7.2 $\pm$0.63 \\    
\object{CFHTWIR-Oph78}  & 16:27:26.23 & -24:19:23.1 & M7.75 & 16.4 & this work & 16.35$\pm$0.05 & 14.36$\pm$0.05 & 13.04$\pm$0.05 & 11.64$\pm$0.06 & 11.14$\pm$0.06 & 10.62$\pm$0.07 &  9.93$\pm$0.09 &  7.79$\pm$0.63 \\   
\object{CFHTWIR-Oph96}  & 16:27:40.84 & -24:29:00.8 & M7.75 & 2.4  & this work	     & 14.60$\pm$0.05 & 13.76$\pm$0.05 & 13.19$\pm$0.05 & 12.51$\pm$0.06 & 12.24$\pm$0.06 & 11.79$\pm$0.07 & 10.96$\pm$0.09 &  8.33$\pm$0.64 \\   
\object{GY92~3}	       & 16:26:21.90 & -24:44:39.8 & M8(M7.5)    & 0.0  & 6,5      & 12.34$\pm$0.02 & 11.48$\pm$0.02 & 10.86$\pm$0.02 &  9.97$\pm$0.06 &  9.57$\pm$0.06 &  9.21$\pm$0.06 &  8.46$\pm$0.09 &  5.82$\pm$0.63 \\   	  
\object{GY92~64}	       & 16:26:32.53 & -24:26:35.4 & M8    & 11.0 & 2	     & 16.36$\pm$0.05 & 14.44$\pm$0.05 & 13.17$\pm$0.05 & 12.15$\pm$0.06 & 11.94$\pm$0.06 & 11.71$\pm$0.08 & 11.77$\pm$0.17 &  ...           \\   
\object{GY92~264}	       & 16:27:26.58 & -24:25:54.4 & M8    & 0.0  & 6	     & 13.00$\pm$0.02 & 12.35$\pm$0.02 & 11.84$\pm$0.02 & 11.09$\pm$0.06 & 10.61$\pm$0.06 & 10.18$\pm$0.07 &  9.32$\pm$0.09 &  6.36$\pm$0.63 \\   	  
\object{CFHTWIR-Oph34}  & 16:26:39.92 & -24:22:33.6 & M8.25 & 9.7  & 7	     & 15.97$\pm$0.05 & 14.53$\pm$0.05 & 13.48$\pm$0.05 & 12.21$\pm$0.07 & 11.66$\pm$0.06 & 11.13$\pm$0.24 & 10.16$\pm$0.40 &  ...           \\   
\object{GY92~11}	       & 16:26:22.27 & -24:24:07.1 & M8.5(M6.5,M8.5)  & 8.0  & 9,2,5      & 16.18$\pm$0.05 & 14.73$\pm$0.05 & 13.75$\pm$0.05 & 12.39$\pm$0.06 & 11.56$\pm$0.06 & 10.79$\pm$0.07 &  9.61$\pm$0.09 &  6.12$\pm$0.64 \\   	  
\object{GY92~141}	       & 16:26:51.28 & -24:32:42.0 & M8.5(M8)  & 0.7  & 1,4      & 15.06$\pm$0.05 & 14.35$\pm$0.05 & 13.86$\pm$0.05 & 13.15$\pm$0.05 & 12.84$\pm$0.06 & 12.33$\pm$0.07 & 11.78$\pm$0.10 &  ...           \\       
\object{GY92~310}	       & 16:27:38.63 & -24:38:39.2 & M8.5(M7,M6)  & 5.7  & 2,3,5    & 13.27$\pm$0.02 & 11.93$\pm$0.02 & 11.08$\pm$0.02 & 10.18$\pm$0.06 &  9.77$\pm$0.06 &  9.34$\pm$0.06 &  8.35$\pm$0.09 &  4.78$\pm$0.63 \\    
\object{SONYC-RhoOph-1} & 16:26:56.33 & -24:42:37.8 & M9    & 5.0  & 10	     & 17.46$\pm$0.05 & 16.44$\pm$0.05 & 15.73$\pm$0.05 & 14.74$\pm$0.06 & 14.46$\pm$0.07 & 14.22$\pm$0.23 & ...            &  ...           \\    
\object{CFHTWIR-Oph77}  & 16:27:25.64 & -24:37:28.6 & M9.75 & 10.0 & this work & 18.23$\pm$0.05 & 16.54$\pm$0.05 & 15.36$\pm$0.05 & 14.16$\pm$0.06 & 13.9 $\pm$0.06 & 13.52$\pm$0.12 & 13.42$\pm$0.20 &  ...           \\   	
\object{CFHTWIR-Oph98}  & 16:27:44.20 & -23:58:52.4 & M9.75 & 3.0  & this work & 17.08$\pm$0.05 & 15.78$\pm$0.05 & 14.98$\pm$0.05 & 13.66$\pm$0.08 & 13.32$\pm$0.07 & 12.98$\pm$0.11 & 12.28$\pm$0.13 &  ...           \\   	
\object{CFHTWIR-Oph9}   & 16:26:03.28 & -24:30:25.9 & L0    & 6.6  & this work & 17.76$\pm$0.05 & 16.33$\pm$0.05 & 15.32$\pm$0.05 & 14.12$\pm$0.07 & 13.63$\pm$0.07 & ... 	   & ...            &  ...           \\    
\object{CFHTWIR-Oph18}  & 16:26:19.41 & -24:27:43.9 & L0    & 7.8  & this work & 18.92$\pm$0.05 & 17.16$\pm$0.05 & 16.04$\pm$0.05 & 14.68$\pm$0.03		 &  ... 	  & ...            & 11.73$\pm$0.16            &  ...           \\   	
\object{CFHTWIR-Oph90}  & 16:27:36.59 & -24:51:36.1 & L0    & 2.4  & this work & 16.83$\pm$0.05 & 15.65$\pm$0.05 & 14.85$\pm$0.05 & 13.86$\pm$0.06 & 13.53$\pm$0.06 & 13.15$\pm$0.08 & 12.32$\pm$0.09 &  8.91$\pm$0.68 \\   	  
\object{CFHTWIR-Oph100} & 16:27:46.54 & -24:05:59.2 & L0    & 8.0  & this work & 17.94$\pm$0.05 & 16.34$\pm$0.05 & 15.26$\pm$0.05 & 14.41$\pm$0.07 & 14.17$\pm$0.08 & 13.96$\pm$0.15 & ...            &  ...           \\        
\object{CFHTWIR-Oph103} & 16:28:10.46 & -24:24:20.4 & L0    & 7.5  & this work & 17.74$\pm$0.05 & 16.15$\pm$0.05 & 15.07$\pm$0.05 & 13.98$\pm$0.06 & 13.68$\pm$0.06 & 13.29$\pm$0.11 & 12.97$\pm$0.14 &  ...           \\   	
\object{CFHTWIR-Oph33}  & 16:26:39.69 & -24:22:06.2 & L4    & 3.6  & this work & 18.16$\pm$0.05 & 16.74$\pm$0.05 & 15.68$\pm$0.05 & 14.35$\pm$0.10 & 14.12$\pm$0.08 & ...            & ...            &  ...           \\     
 \hline                                 
   \label{bd} 
   \end{tabular}
\tablefoottext{a}{Spectral types determined from the following studies: 1.~\citet{Luhman1997}; 2.~\citet{Wilking1999}; 3.~\citet{Luhman1999}; 4.~\citet{Cushing2000}; 5.~\citet{Natta2002}; 6.~\citet{Wilking2005}; 7.~\citet{AlvesdeOliveira2010}; 8.~\citet{McClure2010}; 9.~\citet{Comeron2010}; 10.~\citet{Geers2011}.}
   \end{table}
\end{landscape}

Table~\ref{bd} lists the 29 brown dwarfs and 13 very low-mass stars at the substellar limit members of the cluster with spectroscopic confirmation known to date with coordinates and near-IR photometry, which we adopt in the following sections to study their properties. The \emph{JHK$_{s}$} photometry provided for each member is from the WIRCam/CFHT survey \citep{AlvesdeOliveira2010}, except for members that were saturated or out of the field of view, for which we have listed 2MASS photometry. For each member, the spectral type and A$_{V}$ are shown as compiled from the literature or derived from the numerical spectral fitting in this work. We have mid-IR photometry from the 
\emph{Spitzer} space telescope, taken by the C2D legacy project \citep[\emph{From Cores to Disks}]{Evans2003}, which mapped the $\rho$~Ophiuchi molecular cloud with \emph{Spitzer}'s Infrared Camera \citep[IRAC]{Fazio2004} in the 3.6, 4.5, 5.8, and 8.0~$\mu$m bands over a region of 8.0~deg$^{2}$ and with the Multiband Imaging Camera \citep[MIPS]{Rieke2004} in the 24 and 70~$\mu$m bands over a total of 14.0~deg$^{2}$ \citep{Padgett2008}. The data were retrieved from the C2D point-source catalogues of the final data delivery \citep{Evans2005}, using the NASA/ IPAC Infrared Science Archive. All fluxes were converted to magnitudes using the following zero points: 280.9$\pm$4.1, 179.7$\pm$2.6, 115.0$\pm$1.7, 64.1$\pm$0.94~(Jy), for the 3.6, 4.5, 5.8, and 8.0~$\mu$m IRAC bands, respectively, and 7.17$\pm$0.11~(Jy) for the 24~$\mu$m MIPS band. The \emph{Spitzer} catalogues were merged with the WIRCam or 2MASS catalogue, requiring the closest match to be within 1$\arcsec$. A counterpart was found for all substellar members, which were detected in one or more mid-IR bands. Only detections above 2~$\sigma$ were kept, though all objects have 7~$\sigma$ detections in the [3.6] and [4.5] bands. \\

\subsection{Slopes of the spectral energy distributions}

The disk population of the $\rho$~Ophiuchi cluster has been studied in great detail \citep[e.g.,][]{Natta2002,Andrews2007,Furlan2009,Andrews2010,Ricci2010,Cieza2010,McClure2010}. It is found that, despite the young age of the cluster ($\sim$1Myr), many circumstellar disks in $\rho$~Oph are evolved showing features such as millimeter size dust grains in the outer disk region \citep{Furlan2009,Andrews2010,Ricci2010}, inner opacity holes characteristic of transition disks \citep{Cieza2010}, dust settling, and dust grain processing \citep[e.g.,][]{McClure2010}. In this section, and in light of the newly discovered substellar members in our survey, we use SEDs to classify the evolutionary state of each object. The slope of the SED ($\alpha$$=$$\partial$log($\lambda$F$_{\lambda}$)/$\partial$log($\lambda$)) is used to distinguish YSOs into various classes according to their mid-IR (and millimetre) emitting energy \citep{Lada1984,Adams1987,Lada1987}, representing deeply embedded YSOs \citep[Class 0,][]{Andre1993}, embedded YSOs still surrounded by an envelope (Class I), YSOs possessing a circumstellar disk (Class II), flat-spectrum sources which may represent an intermediate stage between Class I and II \citep[FS,][]{Greene1994}, and sources with a weak or debris disk (Class III).

Following the approach of disk studies using IRAC and MIPS photometry in other young clusters \citep[e.g., Chamaeleon I and Taurus,][]{Luhman2008,Luhman2010}, we computed the spectral slopes across several bands (\emph{K$_{s}$}/[8.0], \emph{K$_{s}$}/[24], [3.6]/[8.0], [3.6]/[24]), which allowed us to derive an SED class even for members which were not detected at 24~$\mu$m. All fluxes were first dereddened by the respective extinction value listed in Table~\ref{bd}, using the extinction law from \citet{Flaherty2007}. We have derived in this manner the spectral slopes for 38 of the low-mass stars at the substellar limit and brown dwarf members.  

\begin{figure*}
   \centering
 \includegraphics[width=\linewidth]{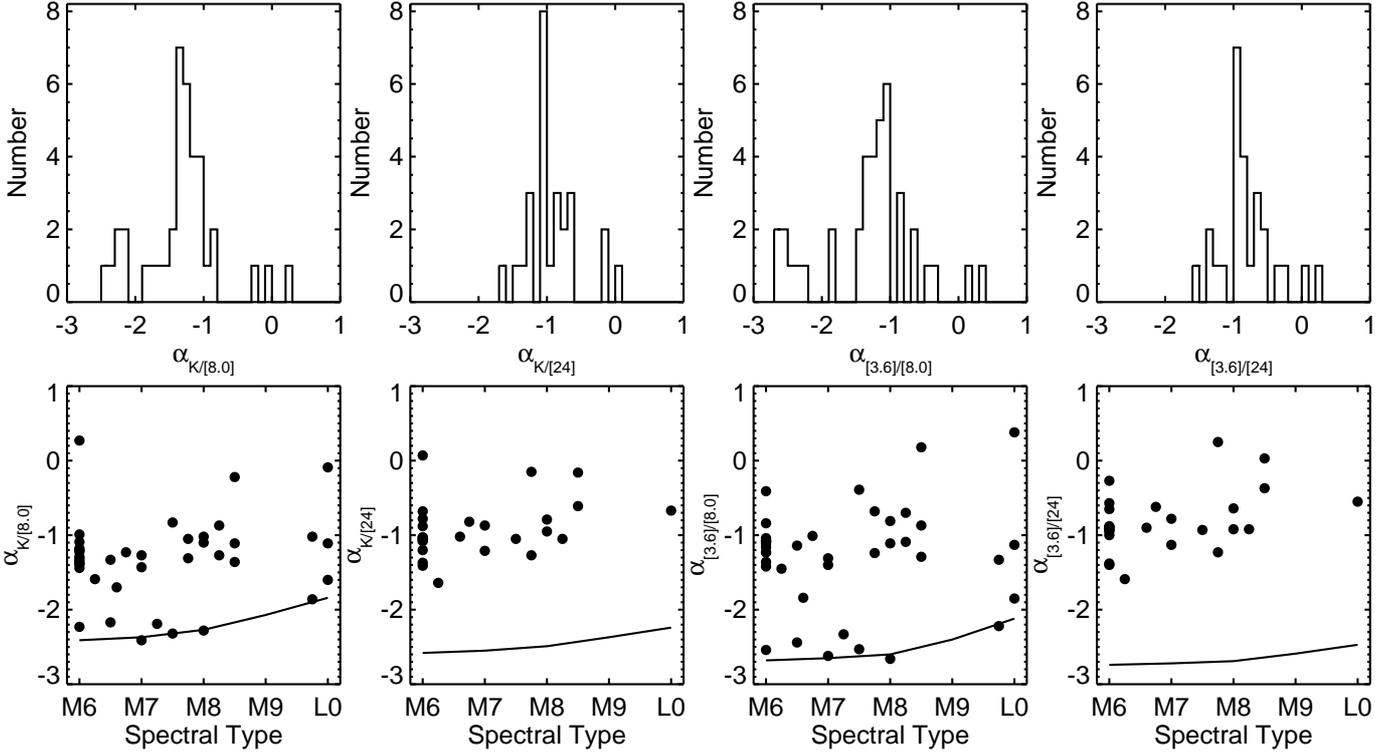}
\caption{SED slopes for spectroscopically confirmed brown dwarfs in $\rho$~Oph. The histograms of the different spectral slopes are shown in the \emph{upper panel}. SED slopes versus spectral type are display in the \emph{lower panel}, where the solid line represents the slopes of the photospheric colours for young objects \protect{\citep{Luhman2010}}.}
\label{slopes}
\end{figure*}

   \begin{table}\label{slopes} 
  \tiny
   \centering             
   \caption{SED slopes for spectroscopically confirmed brown dwarfs in $\rho$~Oph.}           
   \centering             
       \begin{tabular}{l c c c c c}       
   \hline      \hline
ID &  Ks/[8.0] & Ks/[24] & [3.6]/[8.0] & [3.6]/[24] & Class \\ 
   \hline                        
\object{GY92~15}	       & -1.09    & -1.07   & -0.84       & -0.95      &  II     \\      
\object{GY92~109}	       & -1.20    & -1.41   & -1.04       & -1.40      &  II     \\      
\object{GY92~154}	       & -1.18    & -0.68   & -1.23       & -0.57      &  II     \\       
\object{WL21}	       & -1.44    & -1.20   & -1.11       & -1.00      &  II     \\     
\object{GY92~171}	       & -0.99    & -0.88   & -1.16       & -0.92      &  II     \\      
\object{GY92~204}	       & -1.32    & -0.78   & -1.36       & -0.65      &  II     \\    
\object{ISO-Oph~160}        & -1.22    & -1.03   & -1.08       & -0.93      &  II     \\      
\object{GY92~344}	       & 0.27     & 0.07    & -0.41       & -0.27      &  II     \\  
\object{GY92~350}	       & -1.35    & -1.08   & -1.08       & -0.89      &  II     \\     
\object{GY92~371AB}        & -1.39    & -1.37   & -1.42       & -1.38      &  II     \\   
\object{GY92~397}	       & -1.29    & -1.05   & -1.08       & -0.90      &  II     \\     
\object{GY92~450}	       & -2.23    & ...     & -2.54       & ...        &  III     \\        
\object{ISO-Oph~193}        & -1.38    & -1.03   & -1.23       & -0.88      &  II     \\      
\hline
\object{CFHTWIR-Oph~107} & -1.59    & -1.64   & -1.45       & -1.59      &  II     \\        
\object{CFHTWIR-Oph~4}   & -2.17    & ...     & -2.44       & ...        &  III     \\     
\object{CFHTWIR-Oph~106} & -1.33    & ...     & -1.14       & ...        &  II     \\      
\object{CRBR~2322.3-1143}  & -1.70    & -1.02   & -1.84       & -0.90      &  II     \\       
\object{CFHTWIR-Oph~16}  & -1.23    & -0.82   & -1.01       & -0.62      &  II     \\       
\object{GY92~202}	       & -1.27    & -0.87   & -1.31       & -0.78      &  II     \\        
\object{CFHTWIR-Oph~101} & -2.41    & ...     & -2.62       & ...        &  III     \\      
\object{CFHTWIR-Oph~57}  & -2.19    & ...     & -2.33       & ...        &  III     \\      
\object{CRBR~2317.3-1925}         & -0.83    & -1.05   & -0.39       & -0.93      &  II     \\       
\object{CFHTWIR-Oph~47}  & -2.32    & ...     & -2.53       & ...        &  III     \\    
\object{CFHTWIR-Oph~66}  & -1.05    & -0.15   & -0.68       & 0.25       &  II     \\      
\object{CFHTWIR-Oph~78}  & -1.31    & -1.27   & -1.24       & -1.23      &  II     \\        
\object{GY92~3}	       & -1.10    & -0.95   & -1.11       & -0.92      &  II     \\      
\object{GY92~64}	       & -2.28    & ...     & -2.66       & ...        &  III     \\     
\object{GY92~264}	       & -1.02    & -0.79   & -0.81       & -0.64      &  II     \\     
\object{CFHTWIR-Oph~34}  & -0.87    & ...     & -0.70       & ...	       &  II     \\     
\object{CFHTWIR-Oph~96}  & -1.27    & -1.05   & -1.09       & -0.92      &  II     \\      
\object{GY92~11}	       & -0.22    & -0.16   & 0.18        & 0.03       &  II     \\        
\object{GY92~141}	       & -1.36    & ...     & -1.29       & ...        &  II     \\         
\object{GY92~310}	       & -1.11    & -0.61   & -0.87       & -0.37      &  II     \\     
\object{SONYC-RhoOph-1} & ...      & ...     & ...         & ...        &  III?     \\        
\object{CFHTWIR-Oph~77}  & -1.86    & ...     & -2.22       & ...        &  III     \\     
\object{CFHTWIR-Oph~98}  & -1.02    & ...     & -1.33       & ...        &  II     \\     
\object{CFHTWIR-Oph~9}   & ...      & ...     & ...         & ...        &  II?     \\   
\object{CFHTWIR-Oph~18}  & -0.09      & ...     & 0.38         & ...        &  II?     \\       
\object{CFHTWIR-Oph~90}  & -1.11    & -0.67   & -1.13       & -0.55      &  II     \\        
\object{CFHTWIR-Oph~100} & ...      & ...     & ...         & ...        &  III?     \\     
\object{CFHTWIR-Oph~103} & -1.6     & ...     & -1.85       & ...        &  III     \\    
\object{CFHTWIR-Oph~33}  & ...      & ...     & ...         & ...        &  ....     \\   
 \hline                                 
   \end{tabular}
   \end{table}

\subsection{Disk evolutionary states}
Based on the spectral slopes derived, we classify each object into Class~II or Class~III. Figure~\ref{slopes} shows the histograms of the slope values (upper panel) and the distribution of slopes per spectral type (lower panel), which is compared to the slopes of the photospheric colours for young objects \citep[filled line,][]{Luhman2010}. For $\alpha$\emph{(K$_{s}$}/[8.0]) and $\alpha$([3.6]/[8.0]) the histogram shows two well-separated groups corresponding to Class~II ($\gtrsim-$2) and Class~III ($\lesssim-$2) objects. The same groups are also clearly separated in the plot of the slopes as a function of the spectral type, where all objects classified as Class~III follow the line representing stellar photospheres, while the Class II have systematically higher values. The only exception is CFHTWIR-Oph~103, classified as a Class~III since it shows a slope that is only marginally higher then expected for the photospheric colours of a young L0. Due to the limited sensitivity of the 24~$\mu$m measurements, none of the Class III objects has a detection at this wavelength, and therefore does not appear in the $\alpha$\emph{(K$_{s}$}/[24]) and $\alpha$([3.6]/[24]) plots. These diagrams recover, however, all the Class~II. Inside the Class~II group are two objects that have SEDs indicative of Class~I ($\alpha \sim$0), GY92~344 and CFHTWIR-Oph~66. \citep{McClure2010} find evidence for a disk around GY92~344, so we adopt this result and classified it as Class II. CFHTWIR-Oph~66 shows only moderate excess from 3.6 to 8.0~$\mu$m, and is more consistent with a Class~II.

Four objects that are not detected at 8~$\mu$m or 24~$\mu$m (SONYC-RhoOph-1, CFHTWIR-Oph~9, 100, 33) nevertheless have detections at other IRAC bands, which we used to constraint their SED class. CFHTWIR-Oph~100 and SONYC-RhoOph-1 are detected in the three other IRAC channels. In the IRAC colour-colour diagram [3.6]-[4.5]~vs.~[4.5]-[5.8], both objects occupy the same region of the Class~III objects, not showing a significant excess in any of the colours. Based on this evidence, we classified them as likely Class~III (denoted as III?). CFHTWIR-Oph~9 and 33 are only detected at 3.6 and 4.5~$\mu$m. In the colour-colour diagram \emph{K$_{s}$}-[3.6]~vs.~[3.6]-[4.5], CFHTWIR-Oph~9 shows a clear excess mid-IR excess, so we classify it as likely Class~II (denoted as II?). CFHTWIR-Oph~33 occupies a region between Classes~II and III, so we cannot ascertain its SED class. Finally,  CFHTWIR-Oph~18 is detected at 3.6~$\mu$m but has a 2$\sigma$ detection at 8.0~$\mu$m. The slope of the SED shows it has strong mid-IR excess, which places it in the Class~I/Class~II range, though given the limited information at mid-IR wavelengths this classification remains provisory (denoted as II?).

According to the SED classifications, 68$^{+7}_{-10}$\% of the brown dwarf members of $\rho$~Oph are Class~II sources, i.e., have an SED representative of a disk. Statistical errors are calculated following the method proposed in \citet{Burgasser2003}. If we also include all objects at the substellar limit (spectral type $\sim$M6), the percentage of Class~II sources goes up to 76$^{+5}_{-8}$\%. The disk frequency in the stellar domain of $\rho$~Oph has been estimated as 27$\pm5$\% for an extinction-limited sample of A$_{V}$$\lesssim$8 and representative of YSOs with masses $\gtrsim$0.2~$M_{\sun}$ \citep{Erickson2011}. We considered the substellar members that lay within the same extinction limit (see Fig.~\ref{cmd}), and derived a disk fraction of 74\%$^{+7}_{-12}$\%, confirming the large number of brown dwarfs with disks in this cluster. \citet{Luhman2010} have found the disk fraction in Taurus ($\sim$1~Myr) to be 75\% for objects with $\gtrsim$0.3~$M_{\sun}$, and 45\% for low-mass stars and brown dwarfs. By comparing these results to the disk fraction in IC~348, Cha~I, and $\sigma$~Ori, the authors suggest stars with masses greater than 0.5~$M_{\sun}$ members of dense clusters have shorter disk lifetimes. The results of the disk fraction of the stellar population derived by \citet{Erickson2011} also support this hypothesis. The disk fraction we derive for brown dwarfs ($\geqslant$M6) is higher than what is found, for example, in $\sigma$~Ori \citep[60\%,][]{Luhman2008ori}, IC~348 \citep[51\%,][]{Lada2006,Luhman2008}, and Cha~I \citep[44\%,][]{Luhman2008}. A detailed analysis of the disk fraction per spectral type for the $\rho$~Ophiuchi cluster across the entire mass range is needed to understand this result, though such a study is beyond the scope of this paper. 

%__________________________________________________________________

\section{The Rho Ophiuchi population} \label{discussion2}
To compare the properties of the population of the $\rho$~Ophiuchi molecular cloud to other young clusters, it is fundamental to analyse this newly found population of substellar members in respect to the stellar content of the cluster. Despite being one of the closest star-forming regions, the high and variable extinction in $\rho$~Ophiuchi and the fact that the central cluster extends over an area on sky greater than 1~deg$^{2}$ means that many candidate members still lack a spectroscopic classification, hence a reliable membership assessment. Here we summarise the information we gathered from different spectroscopic surveys.

\subsection{The cluster's stellar population} \label{census}
We took as a starting point for the membership compilation the review done on the cluster by \citet{Wilking2008}, excluding the substellar members and members with a spectral type of M6, which are at the substellar limit and have already been discussed in the previous section. To this list, we added newly found members with spectroscopic confirmation published afterwards. The literature review can be summarised as follows:

   \begin{itemize}
      \item \citet{Wilking2008}: in total, the compiled list contains 316 objects, from which 27 are already included in Table~\ref{bd}. From the remaining 289 listed targets, 156 members have assigned spectral types from optical and/or near-IR spectroscopy and evidence of youth (please see the review for references), while 133 did not have any spectral type assigned at that time;
      \item \citet{Cieza2010} derived spectral types for a total of 34 sources, nine of these had previously determined spectral types, while the remaining 25 sources are located outside the central region of the cluster and were therefore not included in our compilation.
      \item \citet{AlvesdeOliveira2010} computed spectral types for six new brown dwarfs and one very low mass star.
      \item \citet{McClure2010} determined spectral types for a few tenths of sources and their binary components (from which 9 are shown in Table~\ref{bd}), including 25 candidate members compiled by \citet{Wilking2008} which lacked spectroscopic and four sources that were not part of that list but have spectral types and membership determined in the literature. N.B., some of these spectral types had been published earlier by \citet{Furlan2009}, though it is in this paper that the data reduction and analysis are presented).
      \item \citet{Geers2011} derived the spectral type for one new brown dwarf.
      \item \citet{Erickson2011} confirmed 30 new members with optical spectral types, where only three were part of the \citet{Wilking2008} compilation. They combined all the new members with 88 previously known members with optical spectral type from an analogous optical survey \citep{Wilking2005}, 14 members compiled from the literature, and three members for which they presented optical spectral types but had already near-IR spectral types determined by \citet{Luhman1999}.
      \item In this work we confirm spectroscopically three new members with spectral types indicative of masses above the substellar limit (besides the 13 brown dwarfs already included in Table~\ref{bd}, and the four members with uncertain spectral types that we do not include in this analysis); 
   \end{itemize}
totalling a list of 250 members of the cluster where 208 have spectral types earlier than M6, and 42 have spectral types later than or equal to M6.

Several other sources have been associated with the cloud  through X-rays \citep{Imanishi2001,Gagne2004,Ozawa2005,Pillitteri2010}, near-IR variability \citep{AlvesdeOliveira2010}, mid-IR excess \citep{Padgett2008,Wilking2008,Gutermuth2009,Marsh2010}, and methane photometric colours \citep{Haisch2010}, but lack spectroscopic confirmation. X-ray surveys have low contamination rates, particularly when large amounts of extinction block background contaminants, though at faint magnitudes, extragalactic contamination is present. In the \citet{Wilking2008} list of candidate members, there are 51 sources that have been associated with the cloud from X-ray surveys but not confirmed spectroscopically. From these, only 11 candidates are within our photometric completeness limits, and more conservative extinction limited sample (see Sect.~\ref{limitedsample}), with \emph{H}-band magnitudes ranging from 9.5 to 14~mag. We therefore do not expect our results to be affected in a significant way. Candidate samples found by the other aforementioned techniques are known to have much higher contamination rates and are therefore not included in the following sections. We note, however, that the majority of these candidates are outside our most conservative extinction limited sample, as well as our completeness limits (see Sect.~\ref{limitedsample}).

While this paper was in the process of revision, \citet{Muzic2011} presented spectra for 21 members of $\rho$~Oph, 14 of which had spectral types already published in the literature and five are simultaneously discovered in this paper (CFHTWIR-Oph~9, 16, 31, 58, 78). The only objects not included in our census are CFHTWIR-Oph~58 for which we could not assign a spectral type (corresponds to SONYC-RhoOph-6 with a spectral type of M6.9 calculated from the H$_{2}$O defined in \citet{Allers2007}), and two low-mass stars (SONYC-RhoOph-3 and 4), which have visual extinctions greater than 8~mag (see Sect.~\ref{limitedsample}). Therefore, we conclude that the results from \citet{Muzic2011} do not affect any of the conclusions of this paper.

\subsection{The H-R diagram}  \label{secthr}
\begin{figure}
   \centering
 \includegraphics[width=\columnwidth]{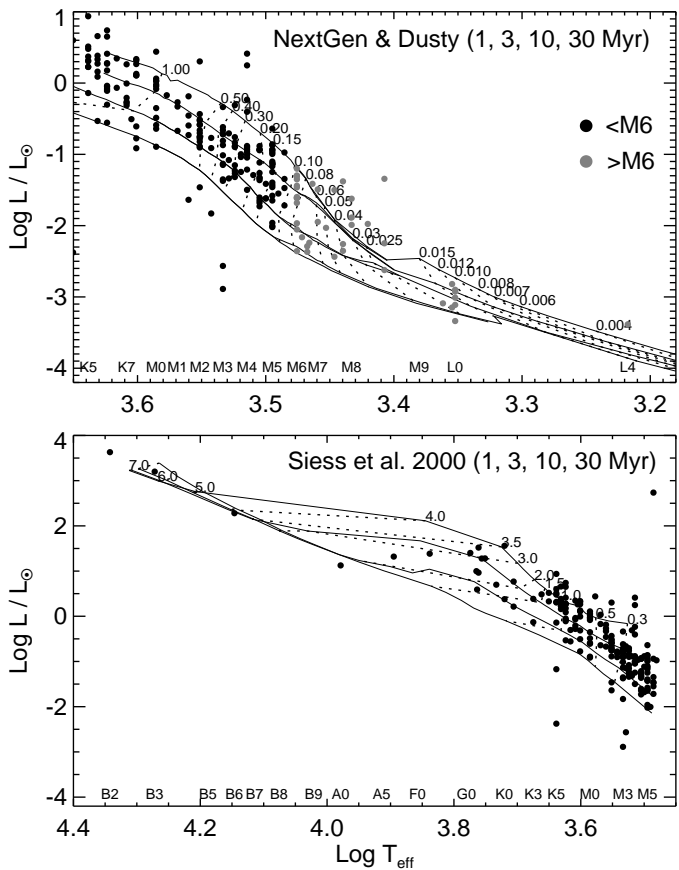}
\caption{H-R diagram for all the spectroscopically confirmed members of $\rho$~Ophiuchi. The models are the NextGen and Dusty isochrones \protect{\citep{Baraffe1998,Chabrier2000}} (upper panel) and the isochrones from \citet{Siess2000} (lower panel) for 1, 5, 10, and 30~Myr, labelled with mass in units of M$_{\sun}$.}
\label{hr}
\end{figure}

To study the masses of the members of the cluster, we first place all the members spectroscopically confirmed in the Hertzsprung-Russell (H-R) diagram (Fig.~\ref{hr}). For the 208 stellar members of the cluster, we compiled information from the literature on the spectral type, A$_{V}$, and near-IR magnitudes. Fourteen objects did not have any measured extinction and are not be displayed in the H-R diagram but are included in the other sections. To convert spectral types to temperature, we adopted the temperature scale from Schmidt-Kahler (1982) for stars earlier than M0, and the scale from \citet{Luhman2003ic348} for sources with spectral type between M0 and M9.5. For the L dwarfs, we applied the scale proposed by \citet{Lodieu2008} extrapolated to the L4 spectral type. The temperatures derived in this way have an estimated uncertainty of $\pm$150~K. The bolometric luminosity was calculated using the dereddened \emph{J} magnitude, a distance to the cloud of 130~pc, and bolometric corrections from Schmidt-Kahler (1982), \citet{Kenyon1995}, and \citet{Dahn2002}. For the most massive members with spectral type B, we plotted the luminosities derived from optical photometry by \citet{Erickson2011}. Figure~\ref{hr} shows all the members overplotted with NextGen and Dusty isochrones \citep{Baraffe1998,Chabrier2000} and the isochrones from \citet{Siess2000}, for 1, 3, 10, and 30~Myr. 

As seen in other young clusters and associations, a large apparent spread in ages is seen in the H-R diagram, with most of the members lying between the 1~Myr and 10~Myrs isochrones, and some objects already close to the 30~Myr isochrone. The age for $\rho$ Oph has been estimated as $\sim$0.3-1 Myr \citep{Greene1995,Luhman1999} by comparing the location on the H-R diagram of near-IR spectroscopically confirmed members found in the central core of the L1688 cloud with pre-main-sequence evolutionary models. The proximity on the sky between the $\rho$~Oph cluster and the southeastern edge of the Upper Sco association (which adjoins the Sco OB2 association), has been several times used to explain the high star-forming activity seen in this cluster. That a supernova explosion occurred $\sim$1.5 Myr ago in Upper Sco is usually invoked to explain a scenario where the strong shock wave ``triggered" star formation in the nearby $\rho$~Ophiuchi \citep{Loren1986,Loren1989}, matching the ages derived for its YSO population. However, this somehow simplistic interpretation should be reviewed in light of more recent theories of cloud formation and lifetime \citep[see, for example, the discussion on Sco-Cen by][]{Preibisch2008}. Based on several observations, one can nevertheless expect that the on-going star formation in $\rho$~Oph is influenced by the nearby high-mass members of Upper Sco \citep[e.g.,][]{Nutter2006}. Observations of a low extinction population, also commonly called the surface population, used optical spectroscopy to confirm many members and by placing then on the H-R diagram derive an older age of 1 to 5~Myr for these objects \citep{Bouvier1992,Martin1998,Wilking2005}. This has been interpreted by \citet{Wilking2005} as an indication that star-formation might have started even earlier in Ophiuchus, simultaneously with Upper Sco. However, there is an increasing scrutiny of the interpretation of the luminosity dispersion seen in the H-R diagram as a true spread in the age of the stars \citep[e.g.,][]{Jeffries2011}, and therefore it could be erroneous to formulate a star formation scenario based only on the luminosity spread. For example, several morphological properties of young stellar objects can affect its derived luminosity, such as the presence of circumstellar emission, variability caused by photospheric spots or the inner disk, and its accretion history. Observationally, unresolved binary systems will also result in an incorrect luminosity, objects observed edge-on, as well as errors in determining extinction, which is non-negligible in this cluster. Finally, errors in the spectral type classification and the fact that we adopt the same distance to all stars will also significantly affect the position in the H-R diagram. In Fig.~\ref{hr}, the few objects located below the 30~Myr evolutionary track have large variability (Oph~30, see Sect.~\ref{var}), while the others are Class~I and Class~II sources, where the circumstellar material is most likely compromising one or more parameters used in deriving their luminosity. 

\subsubsection{Spatially and extinction limited sample} \label{limitedsample}

\begin{figure}
   \centering
 \includegraphics[width=\columnwidth]{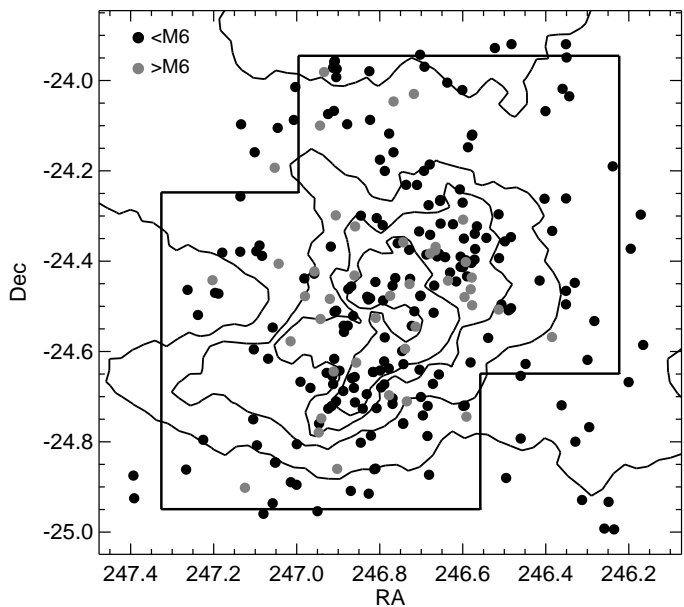}
\caption{Spatial distribution of all the spectroscopically confirmed members of $\rho$~Ophiuchi as compiled from the literature and determined in this paper. The diagram shows the stellar members (black filled circles) and members at the substellar limit and brown dwarfs (grey filled circles), superposed on the contours of the COMPLETE extinction map \citep{Ridge2006}. The field of the WIRCam survey is shown and is used to define a spatially-limited sample. }
\label{sky}
\end{figure}

\begin{figure}
   \centering
 \includegraphics[width=\columnwidth]{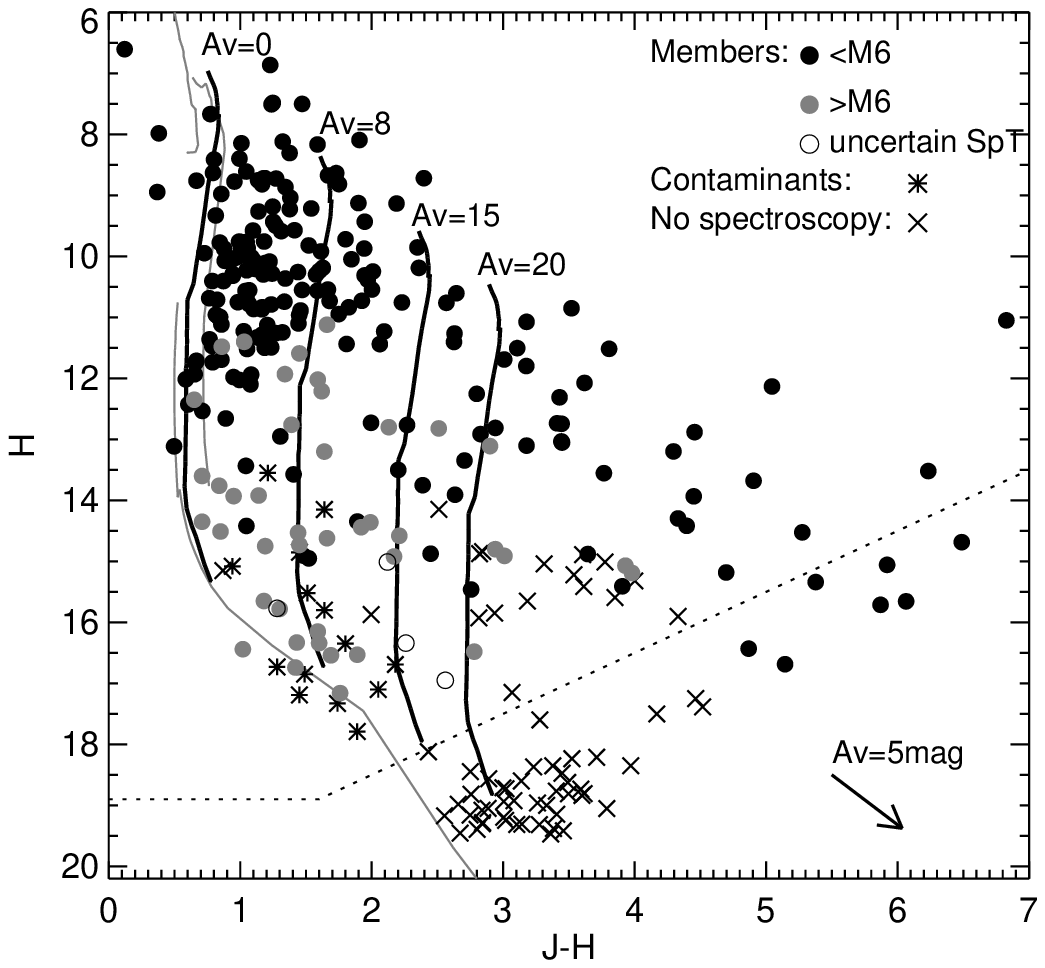}
\caption{Colour-magnitude diagram for the spatially-limited sample of members of $\rho$~Ophiuchi with a known spectral type. The 1~Myr isochrone shifted at a distance of 130~pc from the BT-Settl evolutionary model \citep{Allard2010} is shown for A$_{V}$$=$0, 8, 15, and 20~mag (solid black line), while the 1~Myr isochrones for the NextGen \citep{Baraffe1998}, Dusty \citep{Chabrier2000}, and \citet{Siess2000} are also shown for comparison (solid grey line). The diagram shows the stellar members (black filled circles), members at the substellar limit and brown dwarfs (grey filled circles), members with uncertain spectral types (open circles), contaminants found in our spectroscopic follow-up (asterisks), and candidate brown dwarfs from the WIRCam survey which have not been observed spectroscopically (crosses). The dashed line shows the photometric completeness limits of the WIRCam survey at \emph{J}$=$20.5 and \emph{H}$=$18.9.}
\label{cmd}
\end{figure}

To ensure that the properties we derive for the $\rho$~Ophiuchi cluster are representative of its population, a sample that is spatially and extinction limited, while maximizing completeness must be constructed. In a first step, we select only targets that are contained in the WIRCam survey field (Fig.~\ref{sky}), since we have explored the substellar population of the cluster over this area in an unbiased way. All the selected members are plotted on the colour-magnitude diagram \emph{H} vs. \emph{J}$-$\emph{H} (Fig.~\ref{cmd}), where the completeness limit shown is that of the WIRCam photometric survey \citep{AlvesdeOliveira2010}. In this diagram, we also plot the 1~Myr isochrone at a distance of 130~pc from the BT-Settl evolutionary model \citep{Allard2010}, which ensures a smooth transition between the stellar and substellar regimes. The 1~Myr isochrones for the NextGen \citep{Baraffe1998}, Dusty \citep{Chabrier2000}, and \citet{Siess2000} are also shown for comparison. The isochrone has been reddened by 8, 15, and 20~mag of A$_{V}$, encompassing different depths of the cloud. The extinction-limited sample with A$_{V}$$\lesssim$8 is likely to be complete at all mass ranges, since the combination of optical and near-IR spectroscopic surveys in the stellar regime have targeted these objects. In the substellar regime, we observed all but one candidate brown dwarf (CFHTWIR-Oph~72) spectroscopically down to the completeness limit, defining in this way a nearly complete sample. The limits of A$_{V}$ of 15 and 20~mag are probably not complete. In the substellar regime, the A$_{V}$$\lesssim$15 limit contains two brown dwarf candidates that have not been observed spectroscopically, and this number rises to three if we go to $\lesssim$20. However, the incompleteness is greater in the stellar regime, where at this higher extinction regions, several candidates from X-ray surveys still lack a spectroscopic confirmation and have therefore not been included (see Sect.~\ref{census}). Hereafter, we study the properties of the population with A$_{V}$$\lesssim$8, which we consider to be representative of the cluster, but also display the results for higher extinction regions with the caution that these likely suffer from incompleteness in the stellar regime. 

\subsection{The IMF}
\begin{figure}
   \centering
 \includegraphics[width=\columnwidth]{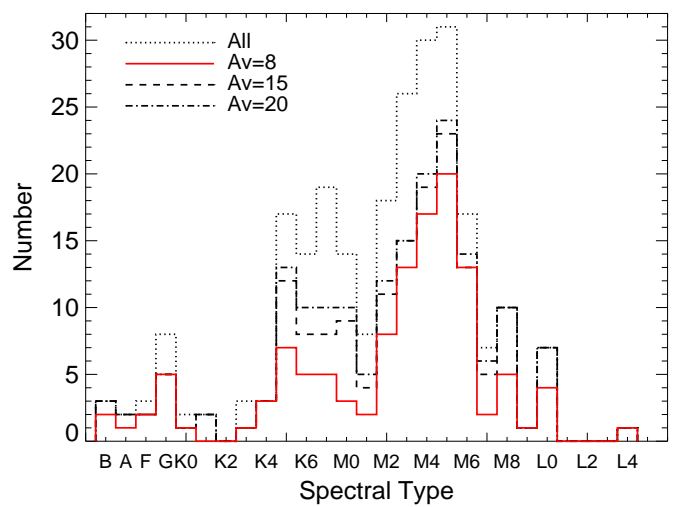}
\caption{Histogram of the spectral types for the population of the $\rho$~Ophiuchi cluster. The lines show the distribution of all members (dotted line), and for the extinction limited samples for A$_{V}$$\lesssim$8 (solid line), 15 (dashed line), and 20~mag (dash-dotted line). The peak of the distribution is located at a spectral type of M5.}
\label{histspt}
\end{figure}

\begin{figure}
   \centering
 \includegraphics[width=\columnwidth]{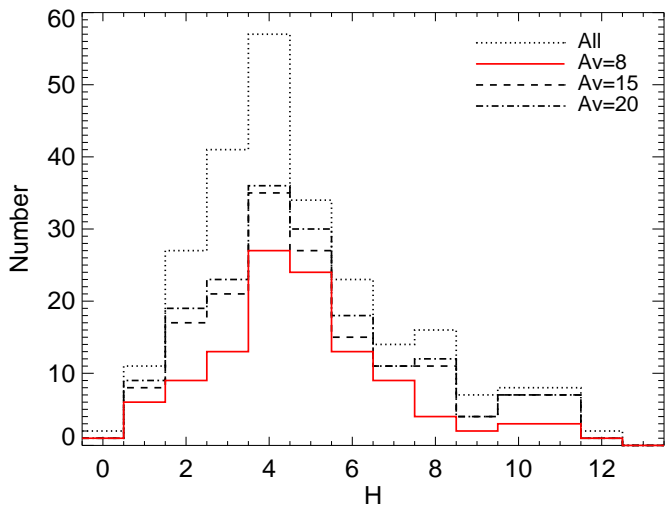}
\caption{Histogram of the absolute \emph{H} magnitudes corrected for extinction and assuming a distance of 130~pc. The distributions are depicted in the same linestyles as Fig.~\ref{histspt}. The histogram peaks at a magnitude of 4.}
\label{histabs}
\end{figure}

\begin{figure}
   \centering
 \includegraphics[width=\columnwidth]{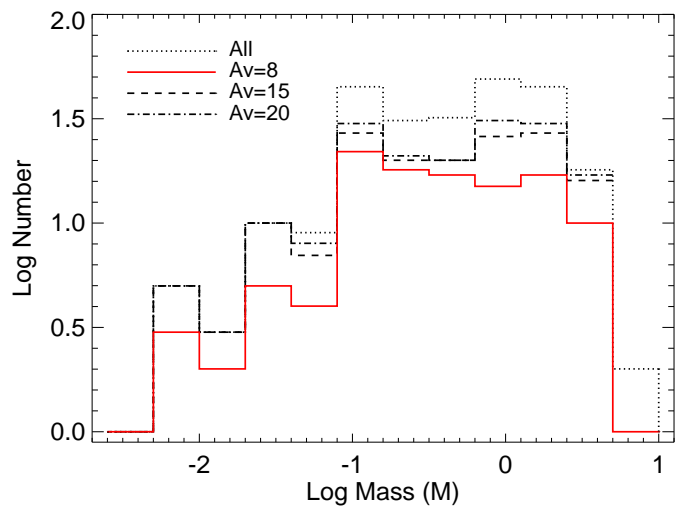}
\caption{IMF for the spectroscopically confirmed population of the $\rho$~Ophiuchi cluster. The peak of the distribution is located at $\sim$0.12~M$_{\sun}$. Linestyles as in Fig.~\ref{histspt}.}
\label{IMF}
\end{figure}

In this section, we analyse the distribution of luminosities and masses of the $\rho$~Oph population and compare it to other star-forming regions. Figure~\ref{histspt} shows the spectral type distribution for our extinction-limited sample. The peak occurs for a spectral type M5 (bin contains members with spectral types $>$4.5 and $\leqslant$5.5), which is similar to the spectral type distributions of other star-forming regions such as IC~348 and Cha~I \citep[e.g.,][]{Luhman2009}. The distribution of the deredden \emph{H} magnitudes corrected for a distance of 130~pc is shown in Fig.~\ref{histabs}, shows a broad peak at an absolute magnitude of 4 to 5. We constructed the IMF for the spectroscopically confirmed members of $\rho$~Oph by deriving masses from the 1~Myr evolutionary models, according to each target's effective temperature. We did not include members with uncertain spectral types, and close binaries that are not resolved in 2MASS or WIRCam images are treated as a single object. The resulting IMF is shown in Fig.~\ref{IMF}, and it exhibits a peak at $\sim$0.12~$M_{\odot}$. This result also agrees with what is found for other young clusters and the field mass function, which have mass function peaking in the 0.1$-$0.2~$M_{\odot}$ range \citep[see, for example, ][for a review]{Bastian2010}.

More quantitatively, we can calculate the number of brown dwarf to stars, defined as $R=N(0.02 \leqslant M/M_{\odot} \leqslant 0.08) / N(0.08 < M/M_{\odot} \leqslant 10)$. This ratio is of 0.09$^{+0.04}_{-0.02}$ for the A$_{V}$$\lesssim$8 sample (and 0.09$^{+0.02}_{-0.02}$ for all members, 0.13$^{+0.03}_{-0.02}$ for A$_{V}$$\lesssim$15, and 0.13$^{+0.03}_{-0.02}$ for A$_{V}$$\lesssim$20). This value is lower than what has been found in other 1-2~Myr old clusters \citep[0.13 in IC~348, 0.18 in Taurus, 0.26 in Cha~I, and 0.44 in NGC~1333,][respectively]{Luhman2003ic348,Luhman2004,Luhman2007cha,Scholz2011}. However, this ratio is highly dependent on the accuracy of the spectral classification at the substellar limit. As pointed out in Sect.\ref{bdcensus}, 13 members of the cluster have been classified with an M6 spectral type through near-IR spectroscopy, a technique that usually carries an error $\pm$1 subclass. According to the evolutionary models used to determine the masses, all these objects have masses greater than $\sim$0.08~M$_{\odot}$. A spectral type of $\sim$M6.25 would, however, place these objects in the substellar regime and significantly change the ratio of brown dwarfs to stars. The value derived for this ratio, if we place the members classified as M6 in the substellar domain, would then be 0.14$^{+0.04}_{-0.03}$ for A$_{V}$$\lesssim$8 (and 0.16$^{+0.03}_{-0.03}$ for all members, 0.19$^{+0.04}_{-0.03}$ for A$_{V}$$\lesssim$15, and 0.20$^{+0.04}_{-0.03}$ for A$_{V}$$\lesssim$20). These values agree with those found for the aforementioned young clusters, with the exception of NGC~1333 claimed to have an excess of brown dwarfs \citep{Scholz2011}. Given that the errors associated with the spectral type classification imply an error in the ratio of brown dwarf to stars which is higher than the purely statistically derived errors, we conclude that within these limitations, our findings agree with other young clusters, and support neither the scenario of a paucity of brown dwarfs in the cluster \citep{Erickson2011}, nor an excess of brown dwarfs down to 0.02~M$_{\odot}$.

%__________________________________________________________________

\section{Conclusions}\label{conclusion}
We have conducted a spectroscopic follow-up of candidate members of $\rho$~Ophiuchi uncovered in our photometric survey. We combine the results of this study with a census of the cluster, which was compiled from the literature to provide an updated study of the YSO population of the cluster. Our findings are summarised as follows.

\begin{itemize}
\item The spectroscopic follow-up resulted in the confirmation of 20 new members of the cluster, from which 13 are brown dwarfs, three are low-mass stars, and four have uncertain spectral types. Combined with our previous spectroscopic follow-up, this amounts to 19 new brown dwarfs members of $\rho$~Oph found through the WIRCam survey. Fourteen of the observed candidates are classified as likely contaminants. Within the photometric completeness limits, we find the success rate of the WIRCam survey to be between 49\% and 74\%.
\item From the new brown dwarfs uncovered in this study, four have a spectral type of L0 and one brown dwarf is an L4. These are the first L-type objects to be found in the cluster, and according to evolutionary models, these objects have masses between $\sim$4 and $\sim$10 Jupiter masses, placing them amongst the least massive young brown dwarfs known to date.
\item We compiled a list of confirmed brown dwarfs in the cluster and using \emph{Spitzer} mid-IR photometry, we derived a ratio of disks of 76$^{+5}_{-8}$\%. This value is higher than what is found in other young clusters, so further studies are needed to understand the implications of these result.
\item A census of the spectroscopically confirmed population of $\rho$~Ophiuchi was compiled from the literature, totalling 250 members. By constructing a spatially and extinction-limited sample, which we believe is representative of the cluster's population, we looked at the luminosity and mass distributions. We find that the distribution of spectral types peaks at M5, similar to what is seen in IC~348 and Cha~I. The mass function peaks at 0.12~M$_{\sun}$, which is consistent with what is found in other clusters and the field. Taking the errors associated with the spectral type classification into account, we find a ratio of brown dwarfs to stars that also in agrees with the one derived for other young clusters. We therefore conclude that there is no evidence for variations in the mass function of $\rho$~Oph when compared to other nearby young clusters.
\end{itemize}

%__________________________________________________________________

\begin{acknowledgements}
We would like to thank Dr. Fernando Selman and Susana Cerda for the excellent assistance during the VLT/ISAAC observations. We thank Dr. Kevin Luhman for kindly providing all young template spectra used in the spectral fitting. We thank the referee for providing helpful comments and suggestions. This publication makes use of data products from the Wide-field Infrared Survey Explorer, which is a joint project of the University of California, Los Angeles, and the Jet Propulsion Laboratory/California Institute of Technology, funded by the National Aeronautics and Space Administration. It is based on observations obtained with WIRCam, a joint project of CFHT, Taiwan, Korea, Canada, France, at the Canada-France-Hawaii Telescope (CFHT), which is operated by the National Research Council (NRC) of Canada, the Institute National des Sciences de l'Univers of the Centre National de la Recherche Scientifique of France, and the University of Hawaii. It is based on observations made with the Gran Telescopio Canarias (GTC), installed at the Spanish Observatorio del Roque de los Muchachos of the Instituto de Astrof'sica de Canarias, on the island of La Palma. This research made use of the SIMBAD database, operated at the CDS, Strasbourg, France. Research was partly supported by the Marie Curie Research Training Network CONSTELLATION under grant no. MRTN-CT- 2006-035890 and by the French National Research Agency under grant 2010 JCJC 0501 1.

\end{acknowledgements}
\bibliographystyle{aa}
\bibliography{ophbib}

\end{document}